\pdfoutput=1
\documentclass[12pt]{article}
\usepackage{amsmath}

\usepackage{amssymb}
\usepackage{graphicx}
\usepackage{adjustbox}
\usepackage{multirow}
\usepackage{booktabs,caption}
\usepackage[flushleft]{threeparttable}
\usepackage{float}
\newcommand{\be}{\begin{equation}}
\newcommand{\ee}{\end{equation}}
\newcommand{\bea}{\begin{eqnarray}}
\newcommand{\eea}{\end{eqnarray}}

\catcode`@=12

\begin{document}
\thispagestyle{empty}
\begin{center}
{\Large\bf
{Signatures of Synchrotron Radiation from the Annihilation of Dark Matter at the Galactic Centre}}\\
\vspace{1cm}
{{\bf Avik Paul} \footnote{email: avik.paul@saha.ac.in},
{\bf Debasish Majumdar} \footnote{email: debasish.majumdar@saha.ac.in}}\\
\vspace{0.25cm}
{\normalsize \it Astroparticle Physics and Cosmology Division,}\\
{\normalsize \it Saha Institute of Nuclear Physics, HBNI} \\
{\normalsize \it 1/AF Bidhannagar, Kolkata 700064, India}\\
{{\bf Amit Dutta Banik} \footnote{email: amitdbanik@iitg.ac.in }}\\
{\normalsize \it Department of Physics ,}\\
{\normalsize \it Indian Institute of Technology, Guwahati 781039, India}\\
\vspace{1cm}
\end{center}
\begin{abstract}
We propose a fermionic dark matter model by extending Standard Model with a Dirac fermion and a real pseudoscalar. The fermion dark matter particle interacts with the Standard Model sector via the Higgs portal through a dimension five interaction term as also through a pseudoscalar interaction term. The parameter space of the model is then constrained by using the vacuum stability and perturbativity condition as also with the LHC constraints. They are finally constrained by the PLANCK results for dark matter relic densities. The direct detection limits are then ensured to have satisfied by the model. We then explore within the framework of the model, the possible signatures of synchrotron radiation from the annihilations of dark matter in the Galactic Centre region  when the end product is $e^+e^-$. We consider the observational data from the radio telescopes namely SKA, GMRT and Jodrell Bank telescopes and compare our calculated synchrotron flux density with them and also with the results predicted by these experiments. We predict that if the low frequency radio telescopes like GMRT, SKA operate at the peak frequencies obtained from our calculations should get a better r.m.s sensitivity.
\end{abstract}
\newpage
\section{Introduction}
The existence of dark matter (DM) in the Universe is now well established from 
various cosmological and astrophysical evidences such as flattening of rotation 
curves of spiral galaxies, gravitational lensing, estimation of total mass and 
its comparison with the luminous mass, the Bullet cluster phenomenon, 
large-scale structures of the Universe, the measurement of anisotropies of 
cosmic microwave background radiation etc. No direct signature of dark matter 
however could be obtained yet by any laboratory experiment and its particle 
nature is still unknown. In the absence of any dark matter signal these 
experiments give lower bounds on dark matter nucleon scattering cross-sections 
for different dark matter masses. Many theoretical dark matter annihilations 
\cite{Ayazi:2015jij}-\cite{Kar:2018rlm} and decay models \cite{Ishiwata:2008qy, 
{Aiba}} are proposed to explain those observations. The indirect signature of dark 
matter may arise out of possible decay and annihilations of dark matter 
particles whereby Standard Model fermion pairs, gamma rays etc. are produced. 
These annihilation products would then give signals in Earth-bound or satellite-borne detectors in excess of those obtained from known astrophysical or 
cosmological processes. The gamma ray excesses observed by Fermi-LAT gamma ray telescope
\cite{Weniger:2012tx, {Bringmann:2012vr}}, the positron excess 
beyond 10 GeV reported by PAMELA \cite{Adriani:2008zr} spaceborne experiment 
and more recently by AMS-02 \cite{AMS-02} experiment on board international 
space station are generally probed for such possible signatures of dark matter indirect detections. 
The satellite-borne experiment DAMPE \cite{Ambrosi:2017wek} also 
reported, from the analysis of their cosmic ray data between 55 GeV to 2.63 TeV, 
an excess of positrons around the energy 1.2 TeV. The terrestrial experiment 
such as HESS \cite{Aharonian:2006au, {Aharonian:2004wa}} and MAGIC 
\cite{Albert:2005kh} also look for TeV gamma ray from the direction of the Galactic Centre (GC)
and if observed an excess then this could be a viable indirect signature for dark 
matter.  

The major content of the dark matter may not be the known fundamental particles 
in the theory of Standard Model of Particle physics. A popular candidate for 
dark matter is WIMPs (Weakly Interacting Massive Particles) 
\cite{Banik:2015aya}-\cite{Biswas:2013nn} 
which would have thermally produced in the early Universe. But depending on their production history in the early Universe they could be of non-thermal types. The dark matter could be very light too and they can be FIMPs (Feebly Interacting Massive Particles) 
\cite{Yaguna:2011qn}-\cite{Molinaro:2014lfa} (generally produced non-thermally), Axions 
\cite{Peccei:2006as}-\cite{Paul:2018msp} or other varieties. Some established and widely discussed theories beyond Standard Model (BSM) such as supersymmetry theory or theories of extra dimensions can predict viable candidates for dark matter such as neutralino \cite{Jungman:1995df},  Kaluza Klein dark matter 
\cite{Bergstrom:2006ny} etc. But these theories are yet to be verified by collider experiments such as Large Hadron Collider (LHC) \cite{Bai:2011wz, 
{Ghosh:2012ep}} or others.
In the present work, we propose a simple 
theoretical model whereby the Standard Model of particle physics is extended by addition of two new extra particles namely a Dirac fermion and a real pseudoscalar for proposing a particle candidate for dark matter. The added Dirac fermion in our model is attributed to a WIMP dark matter particle candidate. In this present model, the dark sector interacts with the SM sector via the Higgs portal through a dimension five interaction term as also through a pseudoscalar interaction term.

In this work we explore the possibility that the dark matter at the Galactic Centre annihilates to produce $e^+e^-$ pairs as the final state particles and under the influence of the Galactic magnetic field emit the synchrotron radiation.
Given that the 
upcoming state of the art radio telescopes such as Square Kilometre Array 
(SKA) \cite{A. vishwas:2010}-\cite{Bertolami:2018lel} as well as the ongoing radio 
telescope experiments such as Giant Metrewave Radio Telescope (GMRT) \cite{Y. 
Gupta:2017,{Y. Guptaet:2018}}, looking for such synchrotron radiations mentioned above could be 
an important signature for indirect searches of dark matter. 

Dark matter can be captured by the gravitational field of very massive astrophysical 
object such as a star, galaxy, galaxy cluster or even a black hole. When 
accumulated in considerable numbers these dark matters may undergo self-annihilations to produce Standard Model (SM) particles such as  electrons 
($e^-$), positrons ($e^+$), muons, neutrinos etc. In case $e^+e^-$ pairs 
generated out of dark matter annihilation at the Galactic Centre region 
then these electrons under the influence of the magnetic field in the vicinity 
of the GC will emit synchrotron radiation. As mentioned earlier, these radiations could be very 
important signals in the detectors like SKA, GMRT etc. In this work we calculate 
the possible flux densities of these synchrotron radiations resulting from possible 
dark matter annihilation in the GC region and its detectability at SKA and 
GMRT. With this in view, as mentioned earlier we propose a particle physics 
model by extending the SM of particle physics with the addition of two new extra 
fields namely a Dirac fermion field and a real pseudoscalar field out of which the Dirac fermion is treated as the dark matter candidate. By imposing 
suitable symmetry we 
establish that the added fermion could be a viable candidate for dark matter in 
our proposed model. We have ensured that our proposed dark matter candidate 
satisfies the PLANCK \cite{Ade:2013zuv} result for the dark matter relic 
density, the collider bound given by the LHC as well as the limits are given by the 
ongoing dark matter direct search experiments such as XENON-1T 
\cite{Aprile:2015uzo}, LUX \cite{Akerib:2016vxi}, PandaX-II \cite{Tan:2016zwf}, XENON-nT \cite{Aprile:2015uzo}, SuperCDMS \cite{Agnese:2015}, DARWIN \cite{Aalbers:2016jon}. With our proposed 
dark matter candidate we calculate their annihilation cross-sections at the GC 
region and obtain the electron-positron spectrum from such annihilations. The frequency response 
of synchrotron radiation caused by these electrons in the magnetic field at the 
GC area and its detectability by SKA and GMRT radio telescopes are then calculated.

The paper is organised as follows. In Section 2, we briefly describe our 
proposed particle physics model of a fermionic dark matter.
In Section 3, we present the constraints on the 
model parameter space from vacuum stability, perturbativity, relic 
density and LHC results. The dark matter phenomenology including 
both the calculations of relic density and direct detection cross-sections
are presented in Section 4. These are used to further constrain the model parameter space by comparing them with the experimental results or bounds.
The viable model parameter space consistent with the 
above-mentioned bounds is also presented in this Section. 
In Section 5, we calculate using our model, the possible signatures of 
synchrotron radiation from the dark matter annihilations. The experimental 
detection range of synchrotron radiation that we have used for our work
along with the results is also given in Section 5. 
Finally in Section 6, we summarise our work with some concluding remarks.
\section{The Model}
In this work the Standard Model (SM) of particle physics is extended by a Dirac fermion field $\chi$ and a real pseudoscalar field $\phi$. The two new extra fields $\chi$ and $\phi$ are singlets under the SM gauge group. In the present scenario the singlet fermion $\chi$ is considered as a dark matter (DM) candidate and it has a global $\rm{U}(1)_{\rm{DM}}$ charge.
The DM candidate $\chi$ interacts with the SM sector through Higgs portal with a dimension 5 interaction term $\left(H^{\dagger} H\right) \bar{\chi}\chi$. Also $\chi$ interacts with the pseudoscalar $\phi$ through a pseudoscalar interaction term (Yukawa type) $\phi\bar{\chi }\gamma^5 \chi$. It is assumed that the interaction Lagrangian $\mathcal{L}_{\text{int}}$ is CP (charge conjugation and parity) invariant. 

The Lagrangian of our proposed model can be written as
\begin{equation}\label{eq:1}
\mathcal{L}=\mathcal{L}_{\text{SM}}+\mathcal{L}_{\text{DM}}+\mathcal{L}_{\phi }+\mathcal{L}_{\text{int}},
\end{equation}
where $\mathcal{L}_{\text{SM}}$ is the SM Lagrangian and the Lagrangian $\mathcal{L}_{\text{DM}}$ for fermionic DM is 
\begin{equation}\label{eq:2}
\mathcal{L}_{\text{DM}}=\bar{\chi } \left(i\gamma^{\mu }\partial_\mu -m\right)\chi.
\end{equation}
The Lagrangian $\mathcal{L}_{\phi }$ for the pseudoscalar boson $\phi$ can be expressed as
\begin{equation}\label{eq:3}
\mathcal{L}_{\phi }=\dfrac{1}{2} \left(\partial_\mu \phi\right)^2-\dfrac{1}{2} m_0^2 \phi^2-\dfrac{\lambda}{24}\phi^4.
\end{equation}
Note that since Lagrangian is CP invariant $\mathcal{L}_{\phi }$ does not contain terms with odd powers of $\phi$ such as the terms involving $\phi$, $\phi^3$ and $\phi H^2$ etc. The interaction Lagrangian $\mathcal{L}_{\text{int}}$ also includes the mutual interaction terms for the scalar $H$ and the pseudoscalar $\phi$.
The form of interaction Lagrangian $\mathcal{L}_{\text{int}}$ is given by
\begin{equation}\label{eq:4}
\mathcal{L}_{\text{int}}=-\dfrac{g_1}{\Lambda}\left(H^{\dagger} H\right) \bar{\chi}\chi-ig\phi\bar{\chi }\gamma^5 \chi-\lambda_1 \phi ^2 H^{\dagger}H-V_H,
\end{equation}
where $\Lambda$ is a high energy scale and $g$ is the dimensionless coupling constant and $V_H$ is written as
\begin{equation}\label{eq:5}
V_H=\mu_H^2 H^{\dagger}H+\lambda_H \left(H^{\dagger} H\right)^2.
\end{equation}
Thus the renormalisable scalar potential has the form
\begin{equation}\label{eq:6}
V=\mu_H^2 H^{\dagger}H+\lambda _H \left(H^{\dagger}H\right)^2+\dfrac{1}{2} m_0^2 \phi^2+\dfrac{\lambda}{24}\phi^4+\lambda_1 H^{\dagger}H \phi ^2.
\end{equation}

After the spontaneous electroweak symmetry breaking the SM Higgs field acquires a non zero vacuum expectation value (VEV), $v_H$ ($v_H\sim 246$ GeV). The pseudoscalar particle also acquires a VEV $v_{\phi }$ due to spontaneous breaking of CP symmetry. After spontaneous symmetry breaking (SSB) we therefore have,\begin{equation}\label{eq:7}
H=\dfrac{1}{\sqrt{2}}\left(
\begin{array}{c}
 0 \\
 v_H+\tilde{H} \\
\end{array}
\right), \phi =v_{\phi }+S,
\end{equation}
where $\tilde{H}$ is an unphysical Higgs and $S$ is an unphysical pseudoscalar.
After spontaneous symmetry breaking the expression of scalar potential (Eq. (\ref{eq:6})) is given by
\begin{equation}\label{eq:8}
\begin{aligned}
V=\dfrac{\mu_H^2}{2}\left(v_H+\tilde{H}\right)^2+\dfrac{\lambda_H}{4} \left(v_H+\tilde{H}\right)^4+\dfrac{m_0^2}{2}\left(v_{\phi}+S\right)^2+\dfrac{\lambda}{24}\left(v_{\phi}+S\right)^4\\+\dfrac{\lambda _1}{2}\left(v_H+\tilde{H}\right)^2 \left(v_{\phi}+S\right)^2. \hspace{7cm}
\end{aligned}
\end{equation}
The interaction term (the last term of the above equation) introduces a mixing between $S$ and $\tilde{H}$. 
From the minimisation conditions
\begin{equation}\label{eq:10}
\dfrac{\partial V}{\partial \tilde{H}}\Big|_{\tilde{H}=S=0}=\dfrac{\partial V}{\partial S}\Big|_{\tilde{H}=S=0}=0,
\end{equation}
one obtains
\begin{equation}\label{eq:11}
m_0^2=-\dfrac{\lambda}{6} v_{\phi }^2-\lambda_1 v_H^2,
\end{equation}
\begin{equation}\label{eq:12}
\mu_H^2=-\lambda_H v_H^2-\lambda_1 v_{\phi }^2.
\end{equation}
The mass matrix $\cal{M}$ in  $\tilde{H}-S$ basis can be constructed by evaluating $\dfrac{\partial ^2V}{\partial S^2}, \dfrac{\partial ^2V}{\partial \tilde{H}^2}$ and $\dfrac{\partial ^2V}{\partial S\partial \tilde{H}}$. The matrix $\cal{M}$ is therefore given as (using Eqs. (\ref{eq:11}) and (\ref{eq:12})),
\begin{equation}\label{eq:13}\hspace{-1.75cm}
{\cal{M}}=\left(
\begin{array}{cc}
 m_{\tilde{H}}^2 & m_{\tilde{H},S}^2\\
 m_{\tilde{H},S}^2 & m_S^2\\
\end{array}
\right)
\end{equation}
\begin{equation}\label{eq:14}\hspace{-6mm}
=\left(
\begin{array}{cc}
 \dfrac{\partial ^2V}{\partial \tilde{H}^2} & \dfrac{\partial ^2V}{\partial S\partial \tilde{H}}\\
 \\
 \dfrac{\partial ^2V}{\partial S\partial \tilde{H}} & \dfrac{\partial ^2V}{\partial S^2}\\
\end{array}
\right)
\end{equation}
\begin{equation}\label{eq:15}
=\left(
\begin{array}{cc}
 2 \lambda _H v_H^2 & 2 \lambda _1 v_{\phi}v_H\\
 2 \lambda _1 v_{\phi}v_H & \dfrac{1}{3}\lambda v_{\phi}^2\\
\end{array}
\right).
\end{equation}
The mass matrix $\cal{M}$ can be diagonalised by a unitary matrix $U$ through a similarity transformation $U^{\dagger}{\cal{M}} U$. The mass eigenstates $h$, $\rho$ in the diagonal basis are connected to $\tilde{H}$ and $S$ through the rotation matrix $U$,
\begin{equation}\label{eq:16}
\left(
\begin{array}{c}
 h \\
 \rho  \\
\end{array}
\right)=U\left(
\begin{array}{c}
 \tilde{H} \\
 S  \\
\end{array}
\right)
=
\left(
\begin{array}{cc}
 \cos\theta & \sin\theta\\
 -\sin\theta & \cos\theta\\
\end{array}
\right)\left(
\begin{array}{c}
 \tilde{H} \\
 S \\
\end{array}
\right),
\end{equation}
which implies
\begin{equation}\label{eq:17}
h=S\hspace{1mm}\sin\theta+\tilde{H}\hspace{1mm}\cos\theta, \hspace{1mm}\rho =S\hspace{1mm}\cos\theta -\tilde{H}\hspace{1mm}\sin\theta,
\end{equation}
where the expression of the mixing angle $\theta$ is given by
\begin{equation}\label{eq:19}
\tan\theta=\dfrac{y}{1+\sqrt{1+y^2}}, {\rm{where}} \hspace{2mm} y=\dfrac{2 m_{\tilde{H},S}^2}{m_{\tilde{H}}^2-m_S^2}.
\end{equation}
One can obtain the expressions of mass eigenvalues of the scalar fields $h$ and $\rho$ are 
\begin{equation}\label{eq:21}
m_{h,\rho }^2=\dfrac{\left(m_{\tilde{H}}^2+m_S^2\right)}{2}\pm \dfrac{\left(m_{\tilde{H}}^2-m_S^2\right)}{2} \sqrt{1+y^2},
\end{equation}
where the + (-) sign corresponds to $h (\rho)$. Here $h$  
is the physical Higgs (SM Higgs) with mass $m_h$=125.09 GeV \cite{Patrignani:2016xqp} and $\rho$ is the physical pseudoscalar of the model with mass $m_{\rho}$. This is to mention here that in this work we consider both the cases when the pseudoscalar $\rho$ is heavier than $h$ ($m_{\rho}>m_{h}$) and when $\rho$ is lighter than $h$ ($m_{\rho}<m_{h}$).
Using Eqs. (\ref{eq:13})-(\ref{eq:21}) one can 
derive the following relations
\begin{equation}\label{eq:22}
\lambda_H=\dfrac{m_{\rho }^2\sin ^2\theta+m_h^2 \cos ^2\theta}{2 v_H^2},
\end{equation}
\begin{equation}\label{eq:23}
\lambda=\dfrac{ m_{\rho }^2\cos ^2\theta+m_h^2 \sin ^2\theta}{v_{\phi }^2/3},
\end{equation}
\begin{equation}\label{eq:24}
\lambda _1=\dfrac{m_{\rho }^2-m_h^2}{4 v_H v_{\phi }}\sin {2\theta}.
\end{equation}
\section{Constraints}
In this section, we will furnish several constraints and bounds on model dependant parameters from theoretical considerations and experimental observations. These are explored in the following. 

\subsection{Theoretical Constraints}
\noindent \underline{\it Vacuum Stability}

In order to obtain a stable vacuum, the scalar potential of our model has to be bounded from below. The quartic terms of the scalar potential can be written as
\begin{equation}\label{eq:25}
V_4=\lambda _H \left(H^{\dagger}H\right)^2+\dfrac{\lambda}{24}\phi^4+\lambda_1 \phi ^2 H^{\dagger}H.
\end{equation}
The stability of the vacuum potential sets the following conditions on the quartic couplings as \cite{Kannike:2012pe}
\begin{equation}\label{eq:26}
\lambda >0,\hspace{1mm}\lambda_H>0,\hspace{1mm}\lambda \lambda_H>6 \lambda_1^2.
\end{equation}

\noindent \underline{\it Perturbativity}

From the perturbativity conditions the bounds are obtained as \cite{Ghorbani:2014qpa}
\begin{equation}\label{eq:27}
\lambda ,\lambda _1,\lambda _H<4 \pi.
\end{equation}
\subsection{Experimental Constraints}
\noindent \underline{\it PLANCK Constraint on Relic Density}

The observed relic density measured by PLANCK is given by \cite{Ade:2013zuv}
\begin{equation}\label{eq:28}
0.1172\leq \Omega _{\text{DM}}{\rm h}^2\leq 0.1226,
\end{equation}
where ${\rm h}$ is the Hubble parameter in units of 100 km 
$\rm{s^{-1}}\rm{Mpc}^{-1}$ and $\Omega _{\text{DM}}$ is the dark matter relic density normalised to the critical density of the Universe. The dark matter component of our model must satisfy 
the above condition for the dark matter relic density. The calculations for the relic 
density is discussed in Section 4.

\noindent \underline{\it Direct Searches of Dark matter}

Dark matter direct detection experiments provide bounds on dark matter nucleon elastic scattering cross-sections for different dark matter masses. We use such bounds obtained from various direct detection experiments namely XENON-1T 
\cite{Aprile:2015uzo}, LUX \cite{Akerib:2016vxi}, PandaX-II \cite{Tan:2016zwf}, XENON-nT \cite{Aprile:2015uzo}, SuperCDMS \cite{Agnese:2015}, DARWIN \cite{Aalbers:2016jon}. We calculate the dark matter nucleon elastic scattering cross-sections that include the model parameters. We compare our calculated results with the constrained model parameters with these experimental bounds and found our calculated cross-sections for different dark matter masses are below the experimental upper bounds for the same. These experimental bounds are utilised to further constrain our model parameter space.
\noindent \underline{\it Collider Constraints}

As mentioned, we considered $h$ is the eigenstate of SM Higgs with mass 
$m_h$=125.09 GeV which has been discovered by Large Hadron Collider (LHC) 
\cite{Bai:2011wz, {Ghosh:2012ep}}. In the present scenario our model is extended 
by adding a fermionic dark matter component and a real pseudoscalar particle. 
These two new extra particles must be affected by the LHC collider physics 
phenomenology. The signal strength of SM like Higgs boson $h$ is defined as 
\begin{equation}\label{eq:30}
R_1=\dfrac{\sigma\left(pp\rightarrow h \right)}{\sigma ^{\text{SM}} \left(pp\rightarrow h\right)}\dfrac{{\rm{Br}} \left(h\rightarrow xx\right)}{{\rm{Br}}^{\text{SM}}\left(h\rightarrow xx\right)},
\end{equation}
where $\sigma\left(pp\rightarrow h \right)$ denotes the produced cross-section 
of Higgs (which will eventually decaying into the particular mode $x$ ($x$=quark, lepton, gauge 
boson, photon etc.)) and $\sigma ^{\text{SM}}\left(pp\rightarrow h\right)$ refer 
the same for the SM Higgs. Whereas ${\rm{Br}} \left(h\rightarrow xx\right)$ 
indicates the decay branching ratio of Higgs decaying into the particular mode 
$x$ ($x \equiv$ quark, lepton or gauge boson) and for the SM case decay branching ratio 
is ${\rm{Br}}^{\text{SM}}\left(h\rightarrow xx\right)$. The branching ratios can be expressed as 
\begin{equation}\label{eq:33}
\text{Br}^{\text{SM}} (h\rightarrow xx)=\frac{\Gamma ^{\text{SM}} (h\rightarrow xx)}{\Gamma ^{\text{SM}}}, \hspace{1mm}\text{Br} \left(h\rightarrow xx\right)=\dfrac{\Gamma  \left(h\rightarrow xx\right)}{\Gamma},
\end{equation}
where $\Gamma ^{\text{SM}} (h\rightarrow xx)$ is the decay width of SM Higgs decaying into final state particles $x$, $\Gamma ^{\text{SM}}$ is the total 
decay width of SM Higgs boson of mass 125.09 GeV, $\Gamma  \left(h\rightarrow 
xx\right)$ is the decay width of $h$ (Eq. (\ref{eq:17})) decaying into final state particles $x$ and 
$\Gamma$ is the total Higgs decay width. 
Expressing the ratios of both the production cross-sections $\left(pp\rightarrow h\right)$ and $\left(pp\rightarrow \rho \right)$ ($\rho$ is the physical pseudoscalar of the model, Eq. (\ref{eq:17})) with the production cross-section $\left(pp\rightarrow {\rm{SM \hspace{1mm}Higgs}}\right)$ can be expressed as
\begin{equation}\label{eq:31}
\dfrac{\sigma \left(pp\rightarrow h\right)}{\sigma ^{\text{SM}} \left(pp\rightarrow h\right)}=\cos ^2\theta, \hspace{1mm}\dfrac{\sigma  \left(pp\rightarrow \rho \right)}{\sigma ^{\text{SM}} \left(pp\rightarrow h\right)}=\sin ^2\theta.
\end{equation}
The decay width
$\Gamma  \left(h\rightarrow xx\right)$ 
can be written as
\begin{equation}\label{eq:35}
\Gamma  \left(h\rightarrow xx\right)=\cos ^2\theta \hspace{1mm}\Gamma ^{\text{SM}} \left(h\rightarrow xx\right).
\end{equation}
Using Eqs. (\ref{eq:30})-(\ref{eq:35}) we obtain
\begin{equation}\label{eq:36}
R_1=\cos ^4\theta\dfrac{\Gamma ^{\text{SM}}}{\Gamma },
\end{equation}
where 
\begin{equation}\label{eq:37}
\Gamma =\cos ^2\theta \hspace{1mm} \Gamma ^{\text{SM}}+\Gamma ^{\text{inv}}.
\end{equation}
In the above, $\Gamma ^{\text{inv}}$ is the invisible Higgs decay width. If the dark 
matter mass satisfies the condition $m_{\chi}\leq m_h/2$ then  
invisible Higgs decay width is denoted by $\Gamma  \left(h\rightarrow \chi  
\bar{\chi }\right)$ and it can be expressed as
\begin{equation}\label{eq:41}
\Gamma\left(h\rightarrow \chi  \bar{\chi }\right)=\dfrac{g^2}{8\pi}\hspace{1mm}\sin ^2\theta \left({{m_h}^2-4{m_{\chi }}^2}\right)^{1/2}+\dfrac{{\Lambda ^{\prime}}^2\cos ^2\theta  {v_H}^2}{8\pi{m_h}^2}\left({m_h}^2-4{m_{\chi }}^2\right)^{3/2},
\end{equation}
where $m_{\chi}$ is the mass of the dark matter particle $\chi$ and from here we 
consider $\dfrac{g_1}{\Lambda }=\Lambda ^\prime$. 

Similarly the signal strength of non SM Higgs boson particle $\rho$ is given by 
\begin{equation}\label{eq:38}
R_2=\dfrac{\sigma\left(pp\rightarrow \rho \right)}{\sigma ^{\text{SM}} \left(pp\rightarrow h\right)}\dfrac{{\rm{Br}}\left(\rho\rightarrow xx\right)}{{\rm{Br}}^{\text{SM}}\left(h\rightarrow xx\right)}
=\sin ^4\theta \hspace{1mm}\dfrac{\Gamma ^{\text{SM}}\left(m_{\rho } \right)}{\Gamma _1},
\end{equation}
where $\sigma\left(pp\rightarrow \rho \right)$ is the production cross-section of $\rho$ and ${\rm{Br}} \left(\rho\rightarrow xx\right)$ is the branching ratio of $\rho$ decaying into final state particles $xx$. $\Gamma ^{\text{SM}}\left(m_{\rho } \right)$ is the total decay width of non SM Higgs boson of mass $m_{\rho}$, $\Gamma_1$ is the total decay width of $\rho$ and it is of the form  
\begin{equation}\label{eq:39}
\Gamma _1=\sin ^2\theta\hspace{1mm}\Gamma ^{\text{SM}}\left(m_{\rho 
}\right)+\Gamma _1^{\text{inv}}+ (\Gamma(\rho\rightarrow hh)~{\rm for} 
~m_{\rho}\ge m_{h}/2),
\end{equation}
where $\Gamma _1^{\text{inv}}$ is the invisible decay width of $\rho$. If the mass of the scalar boson satisfies the condition $m_{\rho}\leq m_{\chi}/2$ then the scalar boson could decay into dark fermion-dark antifermion ($\bar{\chi} \chi$) pair and contribute to the invisible decay width of $\rho$. The expression of decay width of this channel is given by
\begin{equation}\label{eq:44}
 \Gamma_1^{{\rm{inv}}}\left({\rho}\rightarrow \chi  \bar{\chi }\right)=\dfrac{g^2}{8\pi}\hspace{1mm}\cos ^2\theta \left({{m_{\rho}}^2-4{m_{\chi }}^2}\right)^{1/2}+\dfrac{{\Lambda ^{\prime}}^2\sin ^2\theta \hspace{1mm} {v_H}^2}{8\pi{m_{\rho}}^2}\left({m_{\rho}}^2-4{m_{\chi }}^2\right)^{3/2}.
\end{equation}
It is to be noted that for $m_{\rho}\ge m_{h}/2$, an additional contribution to 
total decay width of $\rho$ due to decay into SM scalar have to be 
considered. The expression for $\Gamma(\rho\rightarrow hh)$ is
\begin{equation}\label{eq:95}
\Gamma\left(\rho\rightarrow hh \right)=\dfrac{1}{16\pi m_{\rho}}b^2 
\sqrt{\left(1-\dfrac{4 m_h^2}{m_{\rho}^2}\right)}\,
\end{equation}
where $b$ contributes to the coupling $g_{\rho h h}$ and is given in Eq. (\ref{eq:87}) in the Appendix.
The invisible decay branching ratio of SM scalar can be defined as
\begin{equation}\label{eq:45}
\text{Br}_{\text{inv}}=\frac{\Gamma ^{\text{inv}}}{\Gamma }\,.
\end{equation}
We use the bound $\text{Br}_{\text{inv}}\leq24\%$ \cite{CERN:2016} (for $m_h\ge m_{\chi}/2$) on invisible decay branching ratio of SM Higgs
given by LHC.

It is to be noted that the mixing between SM Higgs doublet and pseudoscalar
could give rise to new self energy corrections to SM gauge boson. This will
provide a strong bound on the scalar mixing from electroweak precision test
measurements. We consider the scalar mixing $\sin\theta\le0.4$ to be consistent
with the precision measurement limits from 
Ref.~\cite{Robens:2015}- \cite{Dupuis:2016} obtained from 
Higgs signal strength measured with LHC run I \cite{Khachatryan:2015, {Aad:2016}} observation. In Ref. \cite{Robens:2015}, a study of extended Higgs sector with singlet scalar has been performed.
The SM Higgs doublet mixes up with singlet resulting two physical scalars.
The bound on mixing angle are based on various theoretical and experimental limits such as 
1) perturbative unitarity, 2) data from electroweak precision test ($S,T, U$) parameters and also
NLO correction to $W$ boson mass, 3) pertubativity of couplings, boundedness and stability of scalar potential,
4) 95\% confidenece level (C.L.) cross-section upper limits from LEP, LHC due to null results on Higgs seraches
which limit signal strength of non-SM scalar boson and 5) consistency of Higgs signal in experiments.
In Ref. \cite{Dupuis:2016} limits from invisible decay of Higgs is also taken into account. Therefore,
in absence of invisible Higgs decay, signal strength of SM Higgs must satisfy 
the condition $R_1\ge0.84$. We use this limit on Higgs signal strength $R_1$ in 
our work. One may mention here that there is no acceptable bounds from LHC for a possible second scalar other than the SM Higgs.

\section{Dark matter phenomenology}
In this section, we briefly present the calculations of dark matter relic 
density and direct detection cross-section and explore the viable 
parameter space of the model.
\subsection{Relic Density}
\begin{figure}
\centering
\includegraphics[width=8cm,height=6cm]{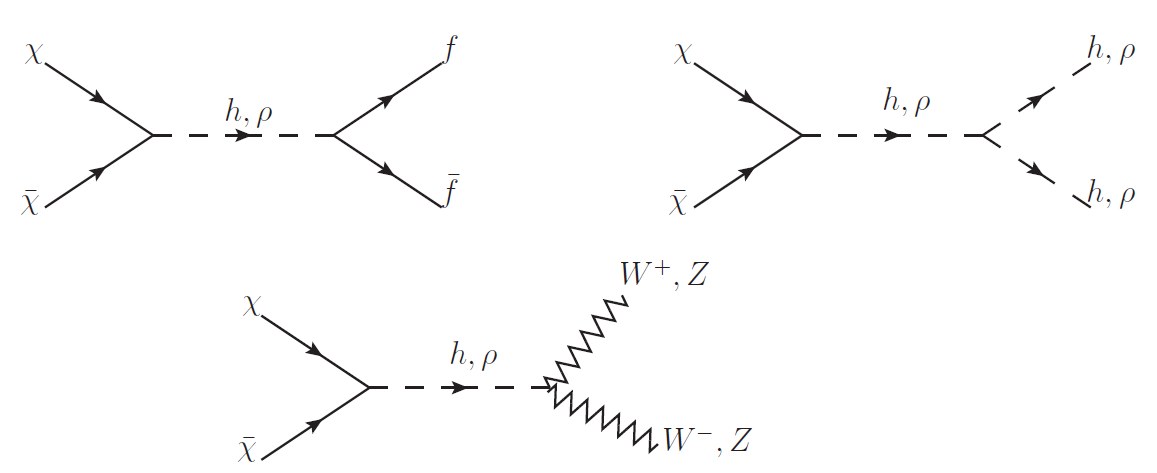}
\caption{Dominant Feynman diagrams for the dark matter annihilation cross-section of the dark matter candidate $\chi$.}
\label{fig:5}
\end{figure}
The relic density of a dark matter candidate $\chi$, is obtained by solving the Boltzmann equation \cite{Kolb:1990vq}
\begin{equation}\label{eq:46}
\frac{dn_{\chi }}{dt}+3Hn_{\chi}=-\langle\sigma v\rangle\left[n_{\chi }^2-\left(n_{\chi }^{\text{eq}}\right){}^2\right],
\end{equation}
where $n_{\chi}$ is the number density of the dark matter candidate $\chi$, $n_{\chi}^{\text{eq}}$ is the number density of the particle $\chi$ at thermal equilibrium and $H$ is the Hubble parameter. The thermal average total annihilation cross-section $\sigma$ times the relative velocity $v$ of the dark matter particle $\langle\sigma v\rangle$ in Eq. (\ref{eq:46}) can be computed by integrating over the centre of mass energy ($\sqrt{s}$) at a temperature $T$ and is given as 
\begin{equation}\label{eq:47}
\left<\sigma v\right>=\dfrac{1}{8m_{\chi}^4 TK_2^2\left(\frac{m_{\chi }}{T}\right)}\int _{4 m_{\chi }^2}^{\infty }ds\left(s-4 m_{\chi }^2\right)\sqrt{s}K_1\left(\frac{\sqrt{s}}{T}\right) \sigma (s),
\end{equation}
where $K_1$ and $K_2$ are the first and second order modified Bessel functions. 
The quantity $\sigma (s)$ as a function of square of the centre of mass energy 
describes the total annihilation cross-section of the dark matter particle 
$\chi$ into final state SM particles (quarks, leptons, gauge bosons, Higgs 
boson) and a pair of scalar bosons $\rho \rho$. The possible $s$ channel 
Feynman diagrams mediated by both $h$ and $\rho$ for the dark matter 
annihilation into the above mentioned final states are demonstrated in Fig. 
(\ref{fig:5}). The expressions for the annihilation cross-sections ($\sigma$) for 
different possible channels relevant for the present dark matter candidate are 
given in the Appendix.
 
The dark matter relic density is then obtained by reducing Eq. (\ref{eq:47}) in 
the form
\begin{equation}\label{eq:56}
\dfrac{1}{Y_0}=\dfrac{1}{Y_F}+\left(\dfrac{45 G}{\pi}\right)^{-\frac{1}{2}} \int_{T_0}^{T_F} g_{*}^{1/2} \langle\sigma {\rm v}\rangle dT,
\end{equation}
where $Y=n_{\chi}/{\rm s}$, ``s'' being the entropy density, $G$ the universal 
gravitational constant, $g_{*}$ the degrees of freedom and $T_F$, $T_0$ are respectively the freeze 
out temperature and the temperature at the present epoch.
Solving for $Y_0$ in Eq.~(\ref{eq:56}), we obtain the expression for dark 
matter relic density as
\begin{equation}\label{eq:57}
\Omega_{\rm DM} {\rm h}^2 =2.755\times 10^8 
\left(\dfrac{m_{\chi}}{\rm{GeV}}\right) Y_0\,.
\end{equation} 
\subsection{Direct detection}
The direct detection of dark matter is realised by the elastic scattering of DM particle with the detector nucleus. As mentioned in Section 2, the fermionic dark matter $\chi$ has interaction terms with both the Higgs $h$ and the pseudoscalar $\rho$. 
The effective Lagrangian for the elastic scattering of DM $\chi$ with the quarks of the nucleus can be expressed as,
\begin{equation}\label{eq:60}
\mathcal{L}_{\text{eff}}=\alpha_{q}\bar{\chi }\gamma^5 
\chi\bar{q}q+\alpha_{q}^{\prime}\bar{\chi }\chi\bar{q}q,
\end{equation}
where $\alpha_{q}=g\sin\theta \cos\theta 
(\frac{m_q}{v_H})\left(\dfrac{1}{m_h^2}-\dfrac{1}{m_{\rho }^2}\right)$, 
$\alpha_{q}^{\prime}=v_H\Lambda^{\prime}\sin\theta \cos\theta 
(\frac{m_q}{v_H})\left(\dfrac{1}{m_h^2}-\dfrac{1}{m_{\rho }^2}\right)$ and $q$ 
denotes the valence quarks. The direct detection scattering cross-section due to the first term (pseudoscalar interaction) of the Lagrangian in Eq. (\ref{eq:60}) is velocity suppressed (by $v_{\rm{rel}}^2$ ($v_{\rm{rel}}\sim 
10^{-3}$)) \cite{Ghorbani:2014qpa, LopezHonorez:2012kv} and hence negligible. The expression for this cross-section is given by
\begin{equation}\label{eq:103}
\left(\sigma_{{\rm{SI}}}\right)_{1}=\dfrac{2g^2}{4\pi } \dfrac{m_r^4}{m_{\chi}^2} \sin 
^2\theta \cos ^2\theta \left(\dfrac{1}{m_h^2}-\dfrac{1}{m_{\rho }^2}\right)^2 
\lambda_p^2 \hspace{1mm}v_{\rm{rel}}^2.
\end{equation}
The dominant contribution to the scattering cross-section is however due to the scalar part (Eq. (\ref{eq:60})) which 
is not velocity suppressed. The spin independent (SI) scattering 
cross-section for the fermionic dark matter $\chi$ in the present model can therefore be approximately written as
\begin{equation}\label{eq:58}
\sigma_{{\rm{SI}}}\backsimeq\dfrac{v_H^2 {\Lambda^{\prime}}^2}{4\pi } m_r^2 \sin 
^2\theta \cos ^2\theta \left(\dfrac{1}{m_h^2}-\dfrac{1}{m_{\rho }^2}\right)^2 
\lambda_p^2,
\end{equation}
where $m_r=\frac{m_N m_{\chi }}{m_N+m_{\chi }}$ is the dark matter nucleon reduced mass with $m_N$ being the mass of the nucleon $N$ and $\lambda_p$ has the following form \cite{LopezHonorez:2012kv}
\begin{equation}\label{eq:59}
\lambda_p=\frac{m_p}{v_H}\left[\sum_q f_q+ \frac{2}{9}\left(1-\sum_q
f_q\right)\right] \simeq1.3\times10^{-3} \,\, .
\end{equation}
With Eqs.~(\ref{eq:58})-(\ref{eq:59}) the direct detection cross-sections of the present dark matter candidate 
are computed and compared with the experimental bounds in order to further constrain the model parameter space.
\subsection{Viable Model parameter space}
In order to obtain the relic density for the fermionic dark matter we first 
calculate the DM annihilation cross-sections (Eqs. (\ref{eq:48})-(\ref{eq:97})) 
into the final state SM particles which are required for calculating the thermal 
averaged annihilation cross-sections (Eq. (\ref{eq:47})). Then we solve the 
Boltzmann equation (Eq. (\ref{eq:56})) to compute $Y_0$ which is necessary for 
the computation of relic density. Finally, the relic densities of the fermionic 
DM for different masses and chosen different sets of parameters are calculated 
using Eq. (\ref{eq:57}). We then constrain the model parameters with the 
theoretical and experimental limits discussed in earlier Section.

\begin{figure}
\includegraphics[width=8.5cm,height=8cm]{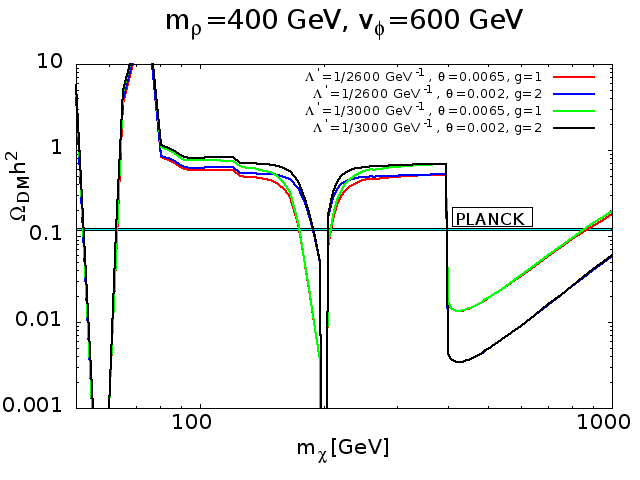}
\includegraphics[width=8.5cm,height=8cm]{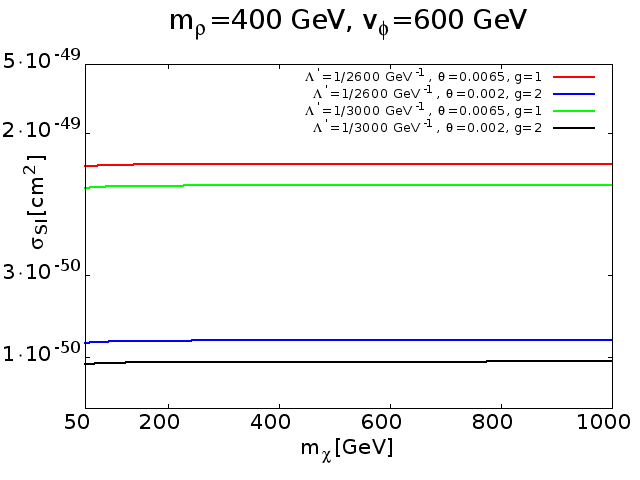}
 \caption{The left panel shows the variation of relic density with the dark matter mass $m_{\chi}$ and the right panel displays the variation of scattering cross-section with the dark matter mass $m_{\chi}$. The graph is plotted with  $m_{\rho}$=400 GeV, $v_{\phi}$=600 GeV and two different choices of $\Lambda^{\prime}$=1/2600 $\rm{GeV}^{-1}$ and 1/3000 $\rm{GeV}^{-1}$.
}
\label{fig:6}
\end{figure}

\begin{figure}
\includegraphics[width=8.5cm,height=8cm]{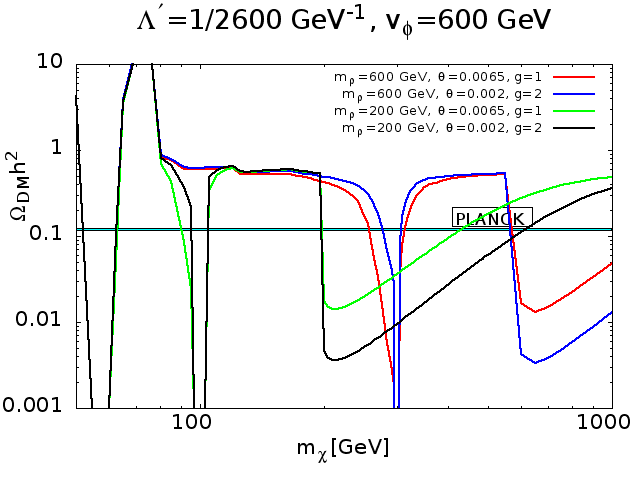}
\includegraphics[width=8.5cm,height=8cm]{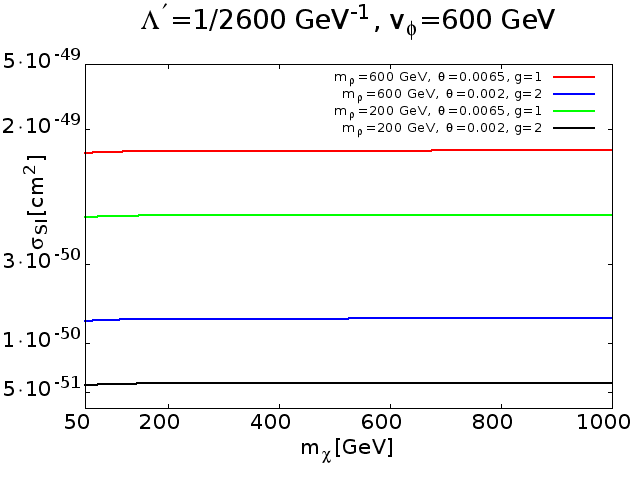}
 \caption{The left panel shows the variation of relic density with the dark matter mass $m_{\chi}$ and the right panel displays the variation of scattering cross-section with the dark matter mass $m_{\chi}$. The graph is plotted with $\Lambda^{\prime}$= 1/2600 $\rm{GeV}^{-1}$, $v_{\phi}$=600 GeV and two different choices of $m_{\rho}$=200 GeV and 600 GeV.}
\label{fig:7}
\end{figure}

\begin{figure}
\includegraphics[width=8.5cm,height=8.5cm]{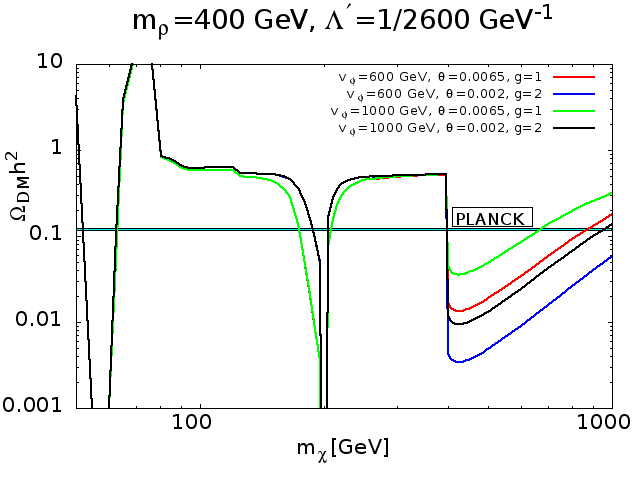}
\includegraphics[width=8.5cm,height=8.5cm]{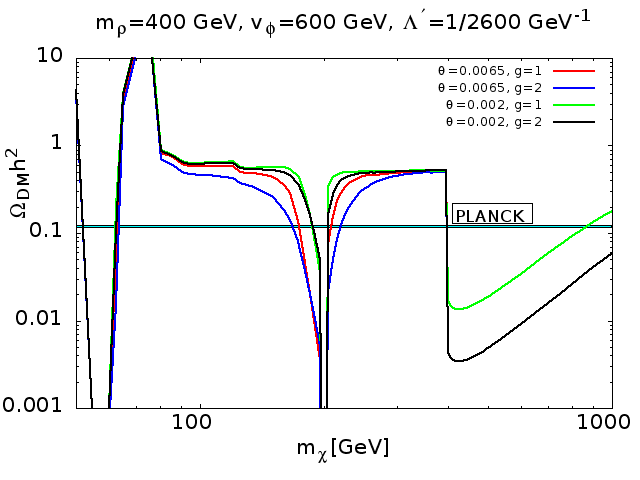}
 \caption{Variation of relic density with the dark matter mass $m_{\chi}$. The left panel is plotted with $\Lambda^{\prime}$=1/2600 $\rm{GeV}^{-1}$, $m_{\rho}$=400 GeV and two different choices of $v_{\phi}$ values 600 GeV, 1000 GeV and for the right panel used $\Lambda^{\prime}$=1/2600 $\rm{GeV}^{-1}$, $m_{\rho}$=400 GeV and $v_{\phi}$=600 GeV.
}
\label{fig:8}
\end{figure}

We calculate the relic density as a function of DM mass between 50 GeV to 1000 
GeV for chosen different values of the parameters. The free parameters ($m_{\rho}, 
\Lambda^{\prime}, v_{\phi}, \theta, g$) are chosen in such a way that they are 
consistent with the vacuum stability, perturbativity and the LHC bounds on 
Higgs signal strength and invisible decay branching ratio. The 
variation of relic density with the DM mass is demonstrated in the left panels of Figs. 
(\ref{fig:6})-(\ref{fig:7}) and both panels of Fig. 
(\ref{fig:8}). We compare the same with the PLANCK observational 
result ($0.1172\leq \Omega_{\text{DM}}{\rm h}^2\leq 0.1226$) which is shown by the
shadowed portion bounded by two parallel lines. 

In Fig.~(\ref{fig:6}) (left panel) we show the variation of dark matter relic 
densities with different dark matter masses for fixed $m_{\rho}$ and $v_{\phi}$. We 
consider two different sets of $g$ and $\theta$ for the plots in Fig.~(\ref{fig:6}) and these are i) $g=1$, $\theta=0.0065$
and ii) $g=2$, $\theta=0.002$. Using the above choice of parameters we plot
$\Omega _{\text{DM}}{\rm h}^2$ with $m_{\chi}$ for two values of 
$\Lambda^{\prime}=1/2600$ $\rm{GeV}^{-1}$, $1/3000$ $\rm{GeV}^{-1}$. The variation of direct detection 
cross-sections $\sigma_{{\rm{SI}}}$ with $m_{\chi}$ for the same set of parameters. From left 
panel of Fig.~(\ref{fig:6}) it can be observed that the relic density of dark 
matter does not vary with variation of $\Lambda^{\prime}$ while in the right panel of Fig.~(\ref{fig:6}) the scattering cross-sections $\sigma_{{\rm{SI}}}$ clearly vary with the change in the value of $\Lambda^{\prime}$. This can be explained in the following way. The contributions to the annihilation cross-sections ($\left<\sigma v\right>$) in principle depend on the both $\Lambda^{\prime}$ and the term arising from pseudoscalar contribution. But the $\Lambda^{\prime}$ dependent term is velocity suppressed (p-wave) and hence the magnitude of the annihilation cross-sections is mostly governed by the pseudoscalar part which is not velocity suppressed. On the other hand from Eq.~(\ref{eq:58}) it is evident that the spin independent scatterring cross-sections $\sigma_{{\rm{SI}}}$ are dependent on $\Lambda^{\prime}$. Also as mentioned earlier, the pesudoscalar part of the scattering cross-section is  velocity suppressed and therefore in  the present work we consider only the scalar part of the dark matter - nucleon scattering cross-sections and its variation with dark matter mass $m_{\chi}$. We observe from Fig.~(\ref{fig:6}) 
(right panel) that direct detection cross-sections for different masses of the dark matter candidate proposed in this work are 
consistent with the latest bounds from different direct search experiments. It may be mentioned that $\Lambda^{\prime}$(=$\frac{g_{1}}{\Lambda})$ is varied by varying the coupling $g_1$ for a chosen high scale value for $\Lambda$ ($>$ TeV). The coupling $g_1$ is varied such that it always remains within the perturbative limit ($g_1<4\pi$). For the variations of $g_1$ within this limit we have checked that our calculated $\sigma_{\rm{SI}}$ for different dark matter masses remain well below the experimental upper bounds for the same including that estimated by the multi-ton scale dark matter experiment namely DARWIN \cite{Aalbers:2016jon}. Direct detection cross-section also depends on the mixing angle $\theta$.
For example for a change of $\theta$ from 0.0065 (set i)) to 0.002 (in set ii)) $\sigma_{{\rm{SI}}}$ is reduced from the value of 1.31$\times 10^{-49}$ to 1.24$\times 10^{-50}$ for $m_{\chi}$=200 GeV.

There are dips in the values of $\Omega _{\text{DM}}{\rm h}^2$ (left panels of Figs. (\ref{fig:6})-(\ref{fig:7}) and both panels of Fig. (\ref{fig:8})) for certain values of $m_{\chi}$.  This phenomenon is attributed to the fact that with the opening up of the new channels, the annihilation cross-sections suffer sudden increases resulting in the decrease of the relic densities. One observes that the relic density decreases for all sets of chosen 
parameters. For example for $m_{\rho}$=400 GeV (left panel of Fig. (\ref{fig:6})) the sudden increase of annihilation cross-section occurs (causing sudden dips in relic density $\Omega _{\text{DM}}{\rm h}^2$) for the DM masses close to 62.5 GeV and 200 GeV. These correspond to resonances at 
$m_h=125.09$ GeV and $m_{\rho}=400$ GeV when dark matter annihilation 
cross-section increases significantly. Also when the DM masses are close to 
$m_{W}$ and $m_{Z}$ two new  DM additional channels open up corresponding to the processes $\bar{\chi } \chi \rightarrow W^+W^-$ and $\bar{\chi } \chi \rightarrow 
ZZ$. In addition, for $m_{\chi}\sim$ mass of the top quark, $\bar{\chi } \chi \rightarrow t \bar{t}$ channel opens up. This channel has also been taken into consideration in our calculations. Similarly one notices dip in relic 
abundance when $\bar{\chi } \chi \rightarrow h h$ channel opens up near 
$m_{\chi}\sim 125$ GeV. Apart from that, for $m_{\chi}\sim  400$ GeV, due to the influence of a new 
annihilation channel ($\bar{\chi } \chi \rightarrow \rho \rho$), the relic abundance suffers another depression. For $m_{\chi}>2m_{\rho}$, with the
increase in DM mass, annihilation cross-section tends to decrease. As a result
DM relic abundance increases and becomes overabundant for higher values of 
$m_{\chi}$. 
In Fig. (\ref{fig:7}) (left panel) we plot the variations of 
$\Omega_{\text{DM}}{\rm h}^2$ with $m_{\chi}$ with $\Lambda^{\prime}$ and 
$v_{\phi}$ are fixed at the values 1/2600 $\rm{GeV}^{-1}$ and 600 GeV 
respectively but for different sets of values for $m_{\rho}=600$ GeV and 200 
GeV using the same set of $\theta, g$ considered in Fig.~(\ref{fig:6}). The 
corresponding plot for direct detection cross-section is shown in the right 
panel of Fig. (\ref{fig:7}).
One can also observe from Fig. (\ref{fig:7}) that changes of $m_{\rho}$ values have significant effects on the relic density plots. When the value of 
$m_{\rho}$ is 200 GeV then the $\bar{\chi } \chi \rightarrow \rho \rho$ channel
opens up at $m_{\chi}$=100 GeV and the contribution of this channel to the annihilation cross-sections dominate after $m_{\chi}$=200 GeV. 
Similarly for $m_{\rho}$=600 GeV the same channel opens up at 300 GeV and the contribution of this channel to the annihilation cross-sections dominates over the other channels with consequent increase of relic density beyond $m_{\chi}$=600 GeV. However, with the increase of dark matter mass, the 
relic density also tends to increase since annihilation cross-section decreases. The latter becomes overabundant as can be seen in Figs.~(\ref{fig:6})-(\ref{fig:7}).
From the right panel of Fig.~(\ref{fig:7}) it is evident that the dark matter direct 
detection cross-sections calculated from our model do not exceed the bounds given by the direct detection experiments. But with the increase of the pseudoscalar mass $m_\rho$, $\sigma_{{\rm{SI}}}$ increases for fixed $g,\theta$ values. This can be 
justified from Eq.~(\ref{eq:58}). For small $m_{\rho}$, the 
factor $\left(\dfrac{1}{m_h^2}-\dfrac{1}{m_{\rho }^2}\right)$ is small but for 
large $m_{\rho}$, $\left(\dfrac{1}{m_h^2}-\dfrac{1}{m_{\rho }^2}\right)\sim 
\dfrac{1}{m_h^2}$. 
However, effects of mixing angle $\theta$ remains same as 
stated earlier in discussions of Fig. (\ref{fig:6}). 
In Fig.~(\ref{fig:8}) (left panel) variations of dark matter relic density with 
dark matter mass are plotted for two representative values of $v_{\phi}$ (VEVs of pseudoscalar) namely $v_{\phi}$=600 GeV and 1000 GeV while the values of $m_{\rho}$ and $\Lambda^{\prime}$ are held fixed. We use the same set of 
$g$ and $\theta$ values considered in Figs.~(\ref{fig:6})-(\ref{fig:7}). We 
observe from the left panel of Fig.~(\ref{fig:8}) that for a fixed set of $g,\theta$, dark matter relic density changes
significantly with $v_{\phi}$ for $m_{\chi}>400$ GeV (i.e., $>2m_{\rho}$).
For $v_{\phi}=600$ GeV dark matter relic density becomes overabundant for 
$m_{\chi}\geq 850$ GeV but for $v_{\phi}=1000$ GeV it reaches required relic 
abundance at 650 GeV. Similar nature is observed for other sets of $g,\theta$. This is due to the fact that with the change of $v_{\phi}$, the triple 
scalar coupling also changes and relic density increases with the increase in 
$v_{\phi}$. Finally in the right panel of Fig.~(\ref{fig:8}) we show how 
dark matter relic density changes with dark matter mass for different sets of parameters
$g$ and $\theta$. Here we have kept other parameters $m_{\rho}, 
\Lambda^{\prime}$ and $v_{\phi}$ fixed. From Fig.~(\ref{fig:8}) (right panel)
we observe that for fixed $g$, changing $\theta$ does not affect significantly the relic density plots. Moreover, for higher values of DM mass 
($m_{\chi}\geq 2m_{\rho}$), the relic densities become invariant of $\theta$ for a fixed values $g$. On the other hand for $m_{\chi}>2m_{\rho}$, the DM relic densities decrease with increase of the parameter $g$ when the value of the parameter $\theta$ is held fixed. 
It is to be noted that direct detection cross-section is independent of
$v_{\phi}$ and $g$. In fact the parameters that $\sigma_{{\rm{SI}}}$  is dependent on are 
$\theta, \Lambda^{\prime}$ and $m_{\rho}$. We have checked that $\sigma_{{\rm{SI}}}$ calculated using the parameters used to generate the plots in Fig.~(\ref{fig:8}), also satisfies the direct detection experimental bounds. Thus we obtained a viable candidate for dark matter by constraining the model parameter space with the dark matter mass ranging from  GeV to TeV. Needless to be mentioned that both the theoretical and experimental bounds are used for constraining the parameters ($g,\theta,\Lambda^{\prime}, v_{\phi}, m_{\rho}$) space. 
\section{Calculation of Synchrotron radiation}
In this section, we estimate the synchrotron radiation signal at terrestrial detectors caused by possible annihilation of dark matter at Galactic Centre. In the present work a fermionic dark matter candidate is considered to undergo the process of self-annihilation in the Galactic Centre region to  produce electrons ($e^-$)and positrons ($e^+$) as the final state particles. Under the influence of the large magnetic field at the GC region these particles are accelerated producing synchrotron radiation. The emitted synchrotron radiation may be detected by radio observations at terrestrial radio telescopes.
\subsection{Formalism}
In order to compute the synchrotron flux for the generated electrons and positrons from dark matter annihilation one needs to solve the standard time-independent diffusion equation which has the following form \cite{Hooper:2012jc}
\begin{equation}\label{eq:64}
K(E)\nabla^{2}n_{e}(E,\textbf{r})+\frac{\partial }{\partial E}\left[b(E,\textbf{r})n_e(E,\textbf{r})\right]+Q(E,\textbf{r})=0,
\end{equation}
where $K(E)$ is the diffusion coefficient as a function of electron energy $E$, $n_{e}(E,\textbf{r})$ is the number density of the electron per unit energy interval at the position $\textbf{r}$, $b(E,\textbf{r})$ is the energy loss coefficient function and $Q(E,\textbf{r})$ is the source term of electrons. 

The produced $e^-$s and $e^+$s lose their energy during their propagation through the galaxy via mainly three processes namely the synchrotron ({\text{synch}}), the inverse Compton ({\text{IC}}) and the bremsstrahlung ({\text{brem}}) processes. The total energy loss rate $b(E,\textbf{r})$ can be expressed as the sum of the energy loss rate components due to these three processes, as
\begin{equation}\label{eq:65}
b(E,\textbf{r})=b_{\text{synch}}(E,\textbf{r})+b_{\text{IC}}(E,\textbf{r})+b_{\text{brem}}(E,\textbf{r}).
\end{equation}

The component $b_{\text{synch}}(E,\textbf{r})$ is given by \cite{Hooper:2012jc}
\begin{equation}\label{eq:66}
b_{\text{synch}}(E,\textbf{r})=\dfrac{dE}{dt}\Big|_{\rm{synch}}=\dfrac{4}{3} \sigma _T c U_{\text{mag}}\left(\textbf{r}\right)\gamma^2\beta^2=3.4\times 10^{-17} \text{GeV} \rm{s}^{-1}\left(\dfrac{E}{\text{GeV}}\right)^2 \left(\dfrac{B}{3\rm{\mu G}}\right)^2,
\end{equation}
where $\sigma _T$ denotes the Thompson scattering cross-section, $B$ is the magnetic field value, $c$ is the velocity of light and $U_{\text{mag}}$ is the magnetic energy density. In Eq. (\ref{eq:66}), $\beta=\sqrt{\gamma^2-1}/\gamma$ where $\gamma=E/m_{e}$, is the  Lorentz factor and $m_{e}$ is the mass of the electron. 
In the present work the calculations are performed with two representative values namely 3 $\mu$G and 6 $\mu$ G for the magnetic field $B$ \cite{Kar:2018rlm, {Ishiwata:2008qy}, {Linden:2010eu}, {Laha:2013}, {Colafrancesco:2015ola}}. 

The energy loss rate by the inverse Compton process has the following form \cite{Hooper:2012jc}
\begin{equation}\label{eq:67}
b_{\text{IC}}(E,\textbf{r})=\dfrac{dE}{dt}\Big|_{\rm{IC}}=\dfrac{2}{9}\dfrac{e^4 U_{\rm{rad}}(\textbf{r})E^2}{\pi \epsilon_0^2 m_e^2c^7}=10^{-16}\text{GeV} \rm{s}^{-1}\left(\dfrac{E}{\text{GeV}}\right)^2\left(\dfrac{ U_\text{rad}(\textbf{r})}{\text{eV}\text{cm}^{-3}}\right),
\end{equation}
where the radiation energy density is denoted by $U_{\rm{rad}}$.

The energy loss rate by the bremsstrahlung process can be estimated as \cite{Springel:2008cc}
\begin{equation}\label{eq:68}
b_{\text{brem}}(E,\textbf{r})=\dfrac{dE}{dt}\Big|_{\rm{brem}}=3\times 10^{-15}\text{GeV} \rm{s}^{-1}\left(\dfrac{E}{\text{GeV}}\right)\left(\dfrac{n_H}{3\hspace{1mm}\text{cm}^{-3}}\right),
\end{equation}
where $n_{H}$ is the number density of hydrogen nuclei in the galaxy.

In Eq. (\ref{eq:64}), the source term of electrons and positrons $Q$ is represented by \cite{Buch:2015iya}
\begin{equation}\label{eq:69}
Q(E,\textbf{r})=\dfrac{1}{2}\left(\dfrac{\rho _{\chi}(\textbf{r})}{m_{\chi}}\right)^2\sum_f\langle\sigma v\rangle_f\dfrac{dN_{e^{\pm}}^f}{dE},
\end{equation}
where $\langle\sigma v\rangle_f$ is the thermal averaged annihilation cross-section of dark matter into final state $f$ and $\frac{dN_{e^{\pm}}^f}{dE}$ is the spectrum of $e^-$s or $e^+$s produced per annihilation of DM into final state $f$. In Eq. (\ref{eq:69}) $\rho _{\chi}(\textbf{r})$ denotes the halo density profile of DM as a function of the galactocentric coordinate $(\textbf{r})$. In this work, we adopt NFW halo profile which is defined as \cite{Navarro:1996gj} 
\begin{equation}\label{eq:71}
\rho_{\text{NFW}}(r)=\dfrac{\rho_s}{\left(\dfrac{r}{r_s}\right)\left(1+\dfrac{r}{r_s}\right)^2},
\end{equation}
where we used $r_s$=20 kpc is the scale radius and the value of the scale density $\rho_{s}$ is chosen in such a way that it can produce the local DM density $\rho_\odot$=0.4 GeV $\rm{cm}^{-3}$ \cite{Catena:2010, {Salucci:2010}} at a distance $r_\odot$=8.5 kpc from Galactic Centre.
In Eq. (\ref{eq:64}), the electron spectrum can be expressed as
\begin{equation}\label{eq:70}
n_e(E,\textbf{r})=\dfrac{\int _E^{m_{\chi }}dE^{\prime}Q\left(E^{\prime},\textbf{r}\right)}{b(E,\textbf{r})}.
\end{equation}
Since the magnetic field $B$ in the GC neighbourhood is very large and the contribution of the energy loss rate (the second term in Eq. (\ref{eq:64})) is significant, the first term in Eq. (\ref{eq:64}) is generally neglected.

The synchrotron power density per unit frequency produced by the electrons and positrons can be written as
\begin{equation}\label{eq:73}
L_{\nu}(\textbf{r})=\int dE\mathcal{P}(\nu ,E)n_e(E,\textbf{r}).
\end{equation}
In the above $\mathcal{P}(\nu,E)$ is expressed as \cite{longair:2010, {lightman:1986}}
\begin{equation}\label{eq:74}
\mathcal{P}(\nu,E)=\dfrac{1}{4 \pi \epsilon_0}\dfrac{\sqrt{3}e^3 B}{m_e c}F\left(\dfrac{\nu}{\nu_c}\right),
\end{equation}
and the critical frequency expressed as \cite{Bertone:2008xr}
\begin{equation}\label{eq:76}
\nu _c=\dfrac{3 e E^2 B}{4\pi m_e^3 c^4}.
\end{equation}
In Eq. (\ref{eq:74}) $F(x)$ takes the form \cite{longair:2010}
\begin{equation}\label{eq:75}
F(x)=x\int _x^{\infty }K_{\frac{5}{3}}\left(x'\right)dx',
\end{equation}
where $K_{\frac{5}{3}}\left(x'\right)$ is the modified Bessel function of order $ \frac{5}{3}$. In earlier works related to synchrotron radiation, the authors \cite{{Bertone:2008xr}, {Regis:2008ij}, {Bertone:2001jv}} considered the approximate formula of $F(x)$ but in this work, we consider the exact form of $F(x)$. 

The synchrotron radiation flux density can be calculated using the expression  
\begin{equation}\label{eq:77}
F_{\nu }=\frac{1}{4 \pi }\int d\Omega \int _{{\rm{l.o.s}}} dl L_{\nu }(\textbf{r}).
\end{equation}
The above integration is performed along the line of sight (l.o.s) distance and over the solid angle $\Omega$. One can determine both these quantities from the following relations 
\begin{equation}\label{eq:72}
r=(r_{\odot}^{2}+l^{2}-2r_{\odot}l\cos{\theta^{\prime}})^{1/2},
\end{equation}
and 
\begin{equation}\label{eq:100}
\Delta \Omega=\int_{\theta^{\prime}_{\rm{min}}}^{\theta^{\prime}_{\rm{max}}} d\theta \sin\theta^{\prime},
\end{equation}
where $\theta^{\prime}$ is the angle between the direction of l.o.s and the line joining Galactic Centre and the Earth, $r$ denotes the distance from the GC to the site in the GC region from  where the annihilation is considered and $r_\odot$ is the distance from the Sun to the Galactic Centre.

We mention here that we have repeated our analysis by adopting one more dark matter density profile namely Einasto profile and furnished our results for two benchmark points namely BP1 and BP3. The Einasto profile is given by \cite{Pieri:2011} 
 \begin{equation}\label{eq:102}
\rho_{\text{Ein}}(r)=\dfrac{\rho_s}{\exp\left[\dfrac{2}{\alpha}\left( \left(\dfrac{r}{r_s} \right)^{\alpha}-1\right)\right]},
\end{equation}
$\alpha=0.17$ for the Einasto profile and the other notations have same significance as in Eq. \eqref{eq:71}.
\subsection{Experimental Detection Range of Synchrotron Radiation Data}
The low frequency radio telescope Giant Metrewave Radio Telescope (GMRT) \cite{Y. Gupta:2017,{Y. Guptaet:2018}} located near Pune in India, has the capability to observe the radio emissions from a wider range astrophysical objects. GMRT is capable of observing the Galactic Centre at an angle $\theta^{\prime}\sim10^{\prime \prime}-1^{\circ}$ \cite{Ghosh:2011, {Roy:2002}} operating in the frequency range of 150-1500 MHz and performs five discrete bands namely 130-170 MHz, 225-245 MHz, 300-360 MHz, 580-600 MHz and 1000-1450 MHz. For the purpose of our present calculations we choose three fixed frequencies namely 325 MHz, 610 MHz and 1400 MHz at observational angles of $10^{\prime \prime}$ and $1^{\prime}$. GMRT observed the flux density of the order of mJy or smaller (1Jy=$10^{-26}\hspace{1mm}{\rm{W\hspace{1mm}Hz}}^{-1}{\hspace{1mm}}{\rm{m}}^{-2}{\hspace{1mm}}{\rm{str}}^{-1}$) and the predicted r.m.s sensitivity in the mJy unit of the above five discrete bands are 0.7, 0.25, 0.04, 0.02 and 0.03 respectively. The upgraded version of GMRT, the uGMRT presently perform in the frequency ranges are 130-260 MHz, 250-500 MHz, 550-900 MHz, 1000-1450 MHz and it will be most sensitive in the low frequency operational band ranges 250 to 1000 MHz \cite{Y. Gupta:2017}. 

A versatile next generation low frequency radio telescope is Square Kilometre Array (SKA) \cite{A. vishwas:2010}-\cite{Bertolami:2018lel} that has been developed for observing the larger area of the sky. Its main objectives to search the origin and evolution of the Universe. SKA operates in the frequency range 70 MHz$-$10 GHz for their observation and our chosen frequencies are well within this range. SKA  predicts upper bounds on the flux density limit which ranges from mJy to $\mu$Jy order.

A new radio telescope MerrKAT \cite{Brederode:2018ghk}, precursor to SKA telescope is under operational which will be most sensitive telescope and will be operated in the three frequency bands namely 0.9-1.67 GHz (L-band), 0.58-1.015 GHz (UHF), 1.75-3.5 GHz (S-band). Its main aim is to investigate cosmic magnetism, Galactic evolution, dark matter, radio sources, large-scale structure etc.

Another important radio telescope, namely Jodrell Bank \cite{{Laha:2013}, {Jodrell}} measures the radio flux in a region of $\theta^{\prime}$=$4 ^{''}$from the GC at frequency 408 MHz with an upper bound on the radio flux of 50 mJy. In the present work, we use the particle dark matter candidate in our model to explore the possibility of detecting the synchrotron radiations at the telescopes mentioned above as indirect signatures for dark matter.
\subsection{Calculations and Results}
In this section, we make a direct comparison between our calculated model 
dependent synchrotron flux density obtained from annihilation of  
dark matter into $e^+e^-$ in final state (produced from direct 
annihilation of DM into other SM particles) with the observational results 
provided by the radio telescopes considered here. The synchrotron flux depends on several factors such 
as dark matter density profile, magnetic field profile in the galaxy, the 
position of the signal in the galaxy, radiation frequency, electron and 
positron spectrum, dark matter mass, corresponding annihilation 
cross-sections etc. In order to obtain the dark matter induced synchrotron flux 
density we first calculate, within the framework of our model, the annihilation cross-sections $\left<\sigma v\right>$ for the dark matter annihilating into $e^+e^-$ and the $e^+e^-$ spectrum. Then using Eq. (\ref{eq:69}) we compute the quantity $Q$. We adopt NFW density 
profile for our work. We then repeat our calculations with the Einasto profile for two benchmark points (BPs) namely BP1 and BP3 and compare our results with those when NFW profile is considered. The calculations are performed by considering two constant magnetic fields of magnitude 3 
$\mu$G and 6 $\mu$G at the GC region and two aperture angles $\theta^{\prime}$=$10^{''}$ and 
$1^{'}$. We like to mention that the choices of the magnetic field values are inspired by the following considerations. In the literatures 
for estimating GC magnetic field, indications are there that this magnetic field would be weak and $\sim$10 $\mu$G \cite{Ferriere:2009dh, {La Rosa:2006}}. In our work we perform the calculations with several values of magnetic field with field strength $\leq$10 $\mu$G and found no significant changes in our results. It may be mentioned that other works in the literature of similar nature \cite{Kar:2018rlm, {Ishiwata:2008qy}, {Linden:2010eu}, {Laha:2013}, {Colafrancesco:2015ola}} also adopted these values for GC magnetic field. The flux density can then be obtained using Eqs. (\ref{eq:65})-(\ref{eq:102}). We use PPPC4DMID 
Mathematica package \cite{{Marco}} for this computation. The calculations are made with six sets of benchmark values (referred to as benchmark points (BPs)) for the parameter sets $m_{\chi}, v_{\phi}, \Lambda^{\prime}, g, m_{\rho}, \theta$. The parameters are so chosen that they satisfy all the theoretical and experimental constraints mentioned earlier. These are furnished in Table \ref{t1}. We also tabulated in Table 1 the relic densities $\Omega_{\text{DM}}{\rm h}^2$ and scattering cross-sections $\sigma_{{\rm{SI}}}$ calculated from each of the six chosen parameter sets (BPs).
\begin{table}[H]
\centering
\begin{tabular}{|l|c|c|c|c|c|c|c|c|r|}
\hline
BP&$m_{\chi}$&$v_{\phi}$&$\Lambda 
^{\prime}$&$g$&$m_{\rho}$&$\theta$&$\Omega_{\text{DM}}{\rm h}^2$&$\sigma_{{\rm{SI}}}$\\
 &in GeV&in GeV&in $\rm{GeV}^{-1}$&&in GeV&&&$\rm{cm^2}$\\
 \hline
1&101.5&600&$\frac{1}{2600}$&1&200&0.0065&0.1187&5.9258$\times 10^{-50}$\\
\hline
2&430&600&$\frac{1}{2600}$&1&200&0.0065&0.1183&6.0151$\times 10^{-50}$\\ 
\hline
3&565&600&$\frac{1}{2600}$&2&600&0.002&0.1195&1.4050$\times 10^{-50}$\\
\hline
4&307&600&$\frac{1}{2600}$&2&600&0.002&0.1187&1.4008$\times 10^{-50}$\\
\hline
 5&131&600&$\frac{1}{2600}$&1&62&0.0065&0.1181&1.5055$\times 10^{-48}$\\
\hline
 6&213&600&$\frac{1}{2600}$&1&100&0.0065&0.1188&5.1012$\times 10^{-50}$\\
\hline
\end{tabular}
\caption{Benchmark points (BPs, six sets) chosen for the present calculations. The relic densities and scattering cross-sections computed using each of the BPs are also given. BP1-BP4 are for the case when $m_{\rho}>m_h$ is chosen and BP5-BP6 are chosen when $m_{\rho}<m_h$. See text for details.}\label{t1}
\end{table}

\begin{figure}[H]
\begin{center}
\includegraphics[width=8cm,height=8cm]{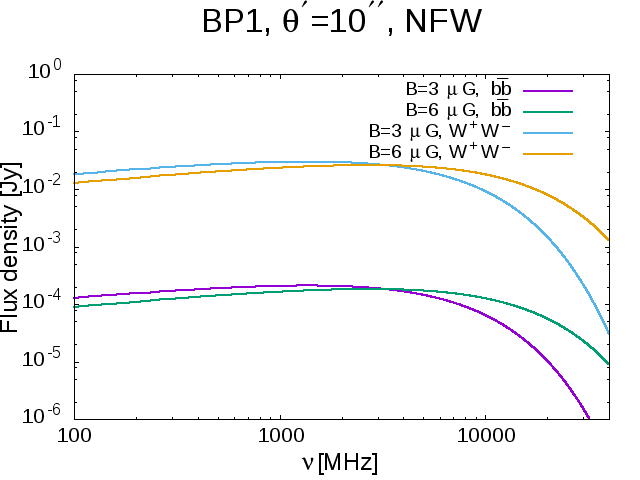}
\includegraphics[width=8cm,height=8cm]{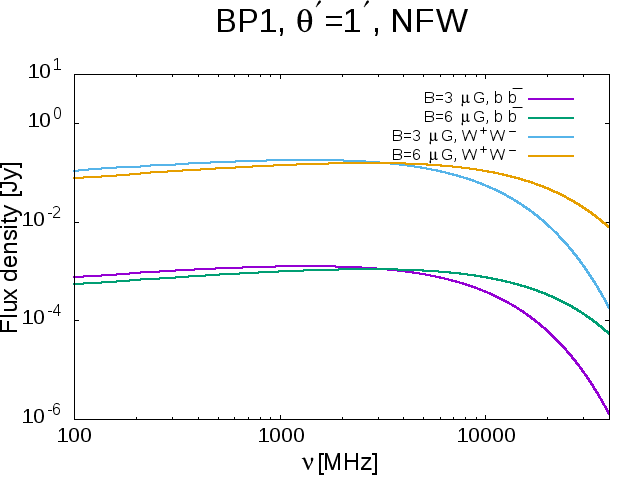}
\caption{Variation of synchrotron flux density with frequency $\nu$ for 
$\theta^{\prime}$=$10^{''}$, $1^{'}$ (from left to right panel) and for BP1 (in Table 
\ref{t1}). In each of the plots, we used two constant magnetic field values namely B=3 
and 6 $\mu$G (from top to bottom).}
\label{fig:9}
\end{center}
\end{figure}

\begin{figure}[H]
\begin{center}
\includegraphics[width=8cm,height=8cm]{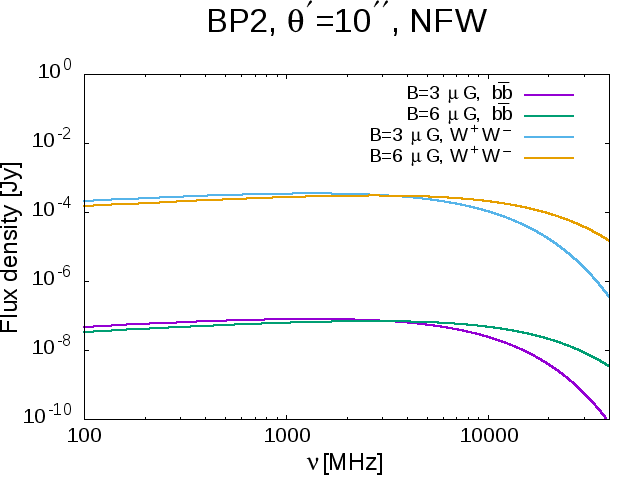}
\includegraphics[width=8cm,height=8cm]{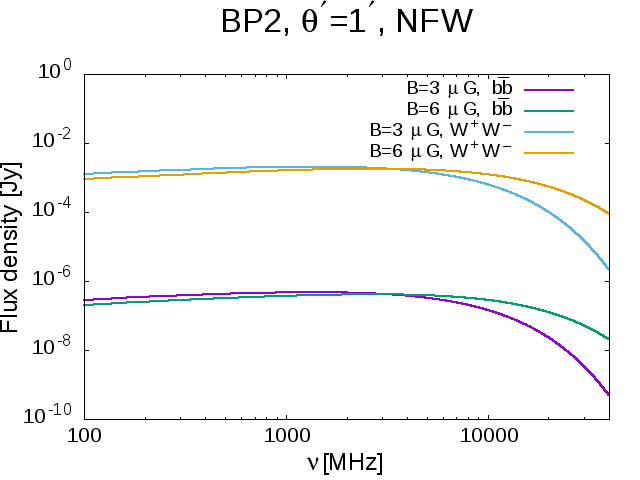}
\caption{Same as in Fig. \ref{fig:9} but for the BP2.}
\label{fig:10}
\end{center}
\end{figure}

\begin{figure}[H]
\begin{center}
\includegraphics[width=8cm,height=8cm]{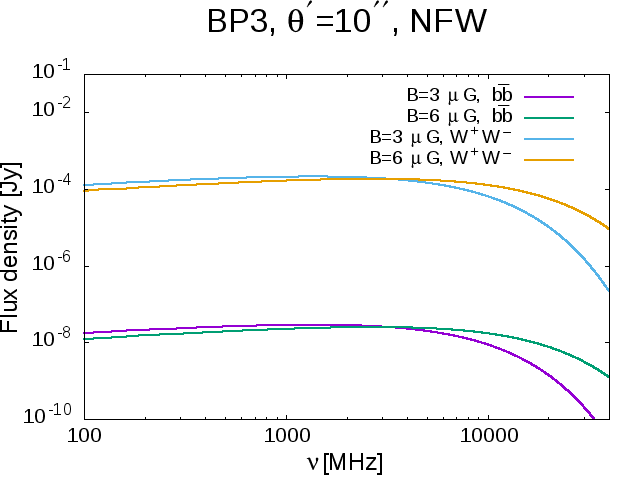}
\includegraphics[width=8cm,height=8cm]{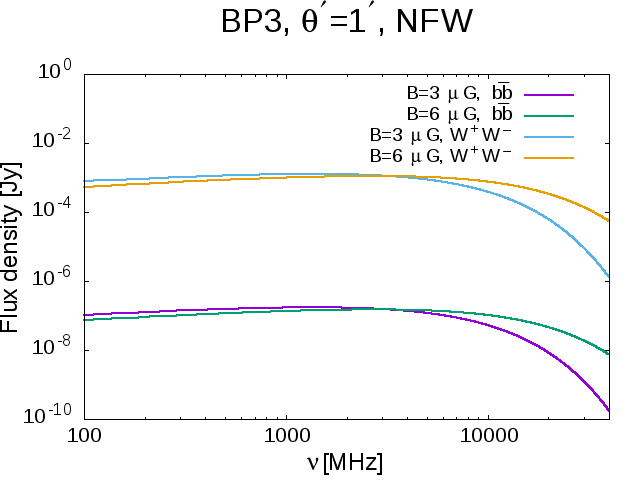}
\caption{Variation of synchrotron flux density with frequency $\nu$ for 
$\theta^{\prime}$=$10^{''}$, $1^{'}$ (from left to right panel) and for the BP3 (in 
Table \ref{t1}). In each of the plots, we used two constant magnetic field values 
namely B=3 and 6 $\mu$G (from top to bottom).}
\label{fig:11}
\end{center}
\end{figure}

\begin{figure}[H]
\begin{center}
\includegraphics[width=8cm,height=8cm]{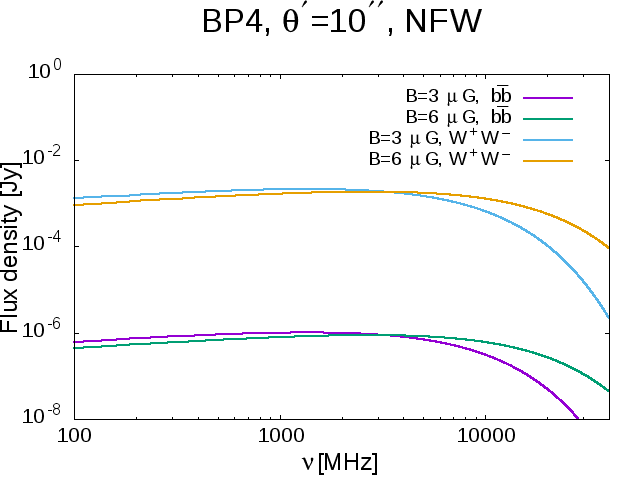}
\includegraphics[width=8cm,height=8cm]{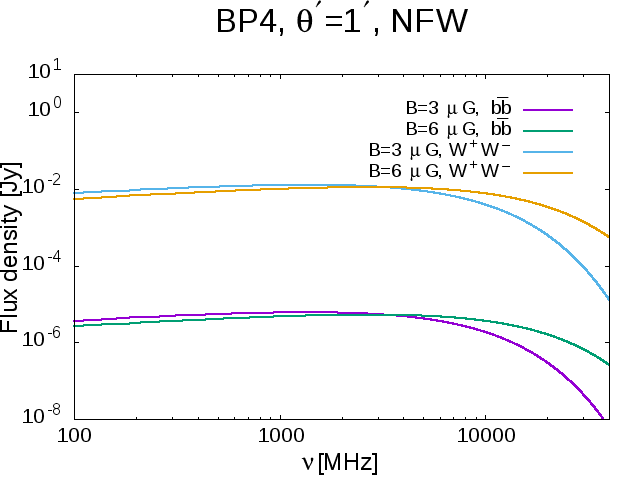}
\caption{Same as in Fig. \ref{fig:11} but for the BP4.}
\label{fig:12}
\end{center}
\end{figure}

In the Figs. (\ref{fig:9})-(\ref{fig:12}) we show the variations of flux 
density with the radiation frequency for chosen four benchmark points (BPs, BP1-BP4) 
which satisfy the relic density and direct detection limits mentioned in 
Table \ref{t1}. We plot the variation of flux density against radiation 
frequency for dark matter primary annihilation completely into $b \bar{b}$ or 
$W^+ W^-$ final state. In Fig. (\ref{fig:9}) we plot the flux density vs 
radiation frequency for $\theta^{\prime}$=$10^{''}$ (left panel) and $1^{'}$ 
(right panel) with BP1. Similar plots are shown in Fig. (\ref{fig:10}) for BP2 
with different dark matter mass $m_{\chi}$=430 GeV.
We also plot radiation frequency against flux density for other two benchmark 
points (BP3 and BP4) depicted in Figs. (\ref{fig:11})-(\ref{fig:12}). 
One can see that from the Figs. (\ref{fig:9})-(\ref{fig:12}) synchrotron flux 
density attains a maximum for every magnetic field values at different 
frequencies. The maximum flux density appears at frequency $\nu$=1.4 GHz for 
$B$=3 $\mu$G and at $\nu$=2.75 GHz for 6 $\mu$G. One can notice from the Figs. 
(\ref{fig:9})-(\ref{fig:12}) the synchrotron flux increases with the angle 
$\theta^{\prime}$. We further observe from Figs. (\ref{fig:9})-(\ref{fig:12})
that at higher frequencies ($\sim$ GHz) flux density decreases sharply for 
lesser values of magnetic field. It is interesting to observe from Figs. 
(\ref{fig:9})-(\ref{fig:12}) that values of flux density obtained for BP1 and 
BP4
are much higher compared to BP2 and BP3. This is due to the fact that BP1 and 
BP4 are near the resonances of the heavy scalar $m_{\chi} \sim m_{\rho}/2$.
Note that in all the above plots, the calculated flux density of dark matter 
annihilating into $W^+W^-$ ($\chi \bar{\chi} \rightarrow W^+ W^-$) channel is 
higher than the flux density obtained from $\chi \bar{\chi}\rightarrow b \bar 
b$ channel. Since dark matter mass is heavier ($m_{\chi}>100$ GeV), $\chi 
\bar{\chi}\rightarrow W^+ W^-$ annihilation dominates over $\chi 
\bar{\chi}\rightarrow b \bar b$ (i.e., $\langle \sigma v \rangle_{\chi 
\bar{\chi}\rightarrow W^+ W^-}>>\langle \sigma v \rangle_{\chi 
\bar{\chi}\rightarrow b \bar b}$). For example,
in case of BP2 with dark matter mass 430 GeV the synchrotron radiation flux 
density for DM annihilation into $b\bar b$ channel is of the order 
$10^{-8}$-$10^{-7}$ Jy while for $W^+W^-$ channel calculated flux density is
about $10^{-4}$-$10^{-3}$ Jy for a large range of frequency (100-1000 MHz). 
We mention here that we have also calculated the variations of flux densities with the radiation frequency for BP5 and BP6 ($m_{\rho}<m_h$ case) and obtained similar variations. However in Table 2 and Table 4 we have furnished the results for these two benchmark points adopted for the case $m_{\rho}<m_h$. 


In Table \ref{t3}, we have calculated synchrotron flux density for six 
different BPs of parameters satisfied by relic density and direct detection 
limits at an aperture angle $\theta^{\prime}$= $10^{''}$ and $1^{'}$ in the GC region and at the 
fixed frequencies namely 325 MHz, 610 MHz and 1400 MHz and then compared it to 
the observations provided by the radio telescopes such as SKA and GMRT. In Table \ref{t3} the results corresponding to BP1 - BP4 are for the case when $m_{\rho}>m_{h}$ and the results for BP5 - BP6 are calculated considering $m_{\rho}<m_{h}$. For all the calculations in Table 2 NFW profile is adopted for dark matter densities.
Note that the three chosen frequencies are in the operational range of GMRT 
and also falls in the operational range of SKA. We consider the 
magnetic field at the region of interest to be $B=3$ $\mu$G and NFW Halo 
density profile of dark matter. Since we observe from the frequency response 
plots for synchrotron radiation flux density (in Figs. 
(\ref{fig:9})-(\ref{fig:12})) that in the operational frequency range of GMRT, 
SKA and Jodrell Bank ($\lesssim $10 GHz) changing the magnetic field from $3\mu$G to $6\mu$G does 
not affect the flux density formidably, we do not expect much change in the 
calculated flux density by changing the value of magnetic field. In fact, we 
expect the results of flux density tabulated in Table \ref{t3} (also in Tables 
\ref{t5}, \ref{t4}, \ref{t6}) will not differ much for the magnetic fields
in the range $B=1\mu$G-10$\mu$G. From Table \ref{t3}, we observe that in case 
of BP1 the flux density calculated for both $b \bar{b}$ and $W^+W^-$ 
annihilation of dark matter exceeds the observational sensitivity from GMRT
for both values of $\theta^{\prime}$= $10^{''}$ and $1^{'}$. However, for other 
benchmark points (BP2 - BP6) although the flux density of synchrotron 
radiation for dark matter annihilating into $b \bar{b}$ is found to be within 
observational limit but for $\chi \bar\chi \rightarrow W^+ W^-$
\begin{table}[H]
\centering 
\footnotesize
\begin{tablenotes}
\centering \small
\item  $B$=3 $\mu$G, NFW density profile.
    \end{tablenotes}
\begin{tabular}{|l|l|l|l|l|l|l|l|}
\hline
BP&\hspace{2mm}$m_{\chi}$& 
\hspace{4mm}$\nu$&\hspace{2mm}$\theta^{\prime}$&\multicolumn{2}{|c|}{
observational limit}&\multicolumn{2}{|c|}{calculated flux 
density}\\\cline{5-6}\cline{7-8}
&&&&\hspace{2mm}\multirow{2}{1mm}{GMRT}  & 
\hspace{4mm}\multirow{2}{1mm}{SKA}&\hspace{8mm}\multirow{2}{1mm}{$\chi 
\bar{\chi}\rightarrow b\bar b$}  & \hspace{4mm}\multirow{2}{1mm}{$\chi 
\bar{\chi} \rightarrow W^+W^-$}\\
&&&&&&&\\
 &in GeV&in MHz&&\hspace{4mm}in Jy&\hspace{4mm}in Jy&\hspace{6mm}in 
Jy&\hspace{4mm}in Jy\\
\hline
1&\multirow{5}{*}{101.5}&325 &\multirow{3}{*}{$10^{''}$} &  0.04$\times 
10^{-3}$& \multirow{5}{*}{$10^{-6}$-$10^{-3}$} &1.7275$\times 
10^{-4}$&2.4706$\times 10^{-2}$\\ \cline{7-8} \cline{3-3} \cline{5-5}
  & &  610   & &  0.02$\times 10^{-3}$ & & 1.9549$\times 10^{-4}$&2.7958$\times 
10^{-2}$\\ \cline{7-8} \cline{3-3} \cline{5-5}
  & &  1400  & &  0.03$\times 10^{-3}$&  & 2.1021$\times 10^{-4}$&3.0064$\times 
10^{-2}$\\ \cline{7-8} \cline{3-3} \cline{4-5}
  & &325&\multirow{3}{*}{$1^{'}$} & 0.04$\times 10^{-3}$ & & 1.0359$\times 
10^{-3}$ &1.4815$\times 10^{-1}$\\ \cline{7-8} \cline{3-3} \cline{5-5}
  & & 610  & &  0.02$\times 10^{-3}$& & 1.1723$\times 10^{-3}$&1.6765$\times 
10^{-1}$\\ \cline{7-8} \cline{3-3} \cline{5-5}
  & & 1400 & &  0.03$\times 10^{-3}$&  & 1.2606$\times 10^{-3}$&1.8028$\times 
10^{-1}$\\ \cline{1-8}
  2&\multirow{5}{*}{430}& 325&\multirow{3}{*}{$10^{''}$} &  0.04$\times 
10^{-3}$& \multirow{5}{*}{$10^{-6}$-$10^{-3}$} 
&$6.5159\times10^{-8}$&2.8442$\times 10^{-4}$ \\ \cline{7-8} \cline{3-3} 
\cline{5-5}
  & &610& &  0.02$\times 10^{-3}$ &  & $7.3736\times10^{-8}$&3.2186$\times 
10^{-4}$\\ \cline{7-8} \cline{3-3} \cline{5-5}
  & &1400 & & 0.03$\times 10^{-3}$&  & $7.9291\times10^{-8}$&3.4611$\times 
10^{-4}$\\ \cline{7-8} \cline{3-3} \cline{4-5}
  & & 325&\multirow{3}{*}{$1^{'}$} & 0.04$\times 10^{-3}$&  &$ 3.9073\times 
10^{-7}$ &1.7056$\times 10^{-3}$\\ \cline{7-8} \cline{3-3} \cline{5-5}
  & & 610& &  0.02$\times 10^{-3}$ &  & $4.4217\times10^{-7}$&1.9301$\times 
10^{-3}$\\ \cline{7-8} \cline{3-3} \cline{5-5}
  & &1400& &    0.03$\times 10^{-3}$ &  & $4.7548\times10^{-7}$&2.0755$\times 
10^{-3}$\\ \cline{1-8}
  3&\multirow{5}{*}{565}&325 &\multirow{3}{*}{$10^{''}$} &  0.04$\times 
10^{-3}$& \multirow{5}{*}{$10^{-6}$-$10^{-3}$} &$2.3737\times 
10^{-8}$&1.7479$\times 10^{-4}$ \\ \cline{7-8} \cline{3-3} \cline{5-5}
  & &610& & 0.02$\times 10^{-3}$ & & $2.6862\times10^{-8}$&1.9780$\times 
10^{-4}$\\ \cline{7-8} \cline{3-3} \cline{5-5}
  & & 14000& & .03$\times 10^{-3}$ & & $2.8885\times10^{-8}$&2.1270$\times 
10^{-4}$\\ \cline{7-8} \cline{3-3} \cline{5-5}
  & & 325&\multirow{3}{*}{$1^{'}$} &0.04$\times 10^{-3}$&  & $1.4234\times 
10^{-7}$ &1.0482$\times 10^{-3}$\\ \cline{7-8} \cline{3-3} \cline{5-5}
  & &  610& & 0.02$\times 10^{-3}$& & $1.6108\times10^{-7}$&1.1861$\times 
10^{-3}$\\ \cline{7-8} \cline{3-3} \cline{5-5}
  & &  1400& & 0.03$\times 10^{-3}$& & $1.7322\times10^{-7}$&1.2755$\times 
10^{-3}$\\ \cline{1-8}
  4&\multirow{5}{*}{307}&325  &\multirow{3}{*}{$10^{''}$} &   0.04$\times 
10^{-3}$& \multirow{5}{*}{$10^{-6}$-$10^{-3}$}  &$8.3656\times 
10^{-7}$&1.7722$\times 10^{-3}$ \\ \cline{7-8} \cline{3-3} \cline{5-5}
  & &610 & &   0.02$\times 10^{-3}$&  & $9.4668\times10^{-7}$&2.0055$\times 
10^{-3}$\\ \cline{7-8} \cline{3-3} \cline{5-5}
  & &1400& &   0.03$\times 10^{-3}$ &  & $1.0180\times10^{-6}$&2.1566$\times 
10^{-3}$\\\cline{7-8} \cline{3-3} \cline{4-5}
  & & 325&\multirow{3}{*}{$1^{'}$} & 0.04$\times 10^{-3}$&  &$ 
5.0165\times10^{-6}$ &1.0627$\times 10^{-2}$\\ \cline{7-8} \cline{3-3} 
\cline{5-5}
  & & 610& &    0.02$\times 10^{-3}$ & & $5.6769\times10^{-6}$&1.2026$\times 
10^{-2}$\\ \cline{7-8} \cline{3-3} \cline{5-5}
  & &1400 & & 0.03$\times 10^{-3}$ & & $6.1045\times10^{-6}$&1.2932$\times 
10^{-2}$\\ \cline{1-8}
 5&\multirow{5}{*}{131}&325 &\multirow{3}{*}{$10^{''}$} &  0.04$\times 
10^{-3}$& \multirow{5}{*}{$10^{-6}$-$10^{-3}$} &7.5272$\times 
10^{-6}$&2.5430$\times 10^{-3}$\\ \cline{7-8} \cline{3-3} \cline{5-5}
  & &  610   & &  0.02$\times 10^{-3}$ & &8.5181$\times 10^{-6}$&2.8778$\times 
10^{-3}$\\ \cline{7-8} \cline{3-3} \cline{5-5}
  & &  1400  & &  0.03$\times 10^{-3}$&  & 9.1597$\times 10^{-6}$&3.0946$\times 
10^{-3}$\\ \cline{7-8} \cline{3-3} \cline{4-5}
  & &325&\multirow{3}{*}{$1^{'}$} & 0.04$\times 10^{-3}$ & &4.5138$\times 
10^{-5}$ &1.5250$\times 10^{-2}$\\ \cline{7-8} \cline{3-3} \cline{5-5}
  & & 610  & &  0.02$\times 10^{-3}$& & 5.1080$\times 10^{-5}$&1.7257$\times 
10^{-2}$\\ \cline{7-8} \cline{3-3} \cline{5-5}
  & & 1400 & &  0.03$\times 10^{-3}$&  & 5.4928$\times 10^{-5}$&1.8557$\times 
10^{-2}$\\ \cline{1-8}
 6&\multirow{5}{*}{213}&325 &\multirow{3}{*}{$10^{''}$} &  0.04$\times 
10^{-3}$& \multirow{5}{*}{$10^{-6}$-$10^{-3}$} & 9.5337$\times 
10^{-7}$&1.0181$\times 10^{-3}$\\ \cline{7-8} \cline{3-3} \cline{5-5}
  & &  610   & &  0.02$\times 10^{-3}$ & & 1.0789$\times 10^{-6}$& 1.1521$\times 
10^{-3}$\\ \cline{7-8} \cline{3-3} \cline{5-5}
  & &  1400  & &  0.03$\times 10^{-3}$&  & 1.1601$\times 10^{-6}$&1.2389$\times 
10^{-3}$\\ \cline{7-8} \cline{3-3} \cline{4-5}
  & &325&\multirow{3}{*}{$1^{'}$} & 0.04$\times 10^{-3}$ & & 5.7170$\times 
10^{-6}$ &6.1052$\times 10^{-3}$\\ \cline{7-8} \cline{3-3} \cline{5-5}
  & & 610  & &  0.02$\times 10^{-3}$& & 6.4696$\times 10^{-6}$& 6.9088$\times 
10^{-3}$\\ \cline{7-8} \cline{3-3} \cline{5-5}
  & & 1400 & &  0.03$\times 10^{-3}$&  & 6.9570$\times 10^{-6}$&7.4293$\times 
10^{-3}$\\ \cline{1-8}
\end{tabular}
 \captionsetup{font=small}
\caption{Model dependent theoretical flux densities for chosen six BPs are 
compared with the observational data are given by GMRT and SKA.}\label{t3}
\end{table}
\hspace{-.8cm} channel it 
exceeds the limit from GMRT (for both values of $\theta^{\prime}$). We further
compare our results with the SKA sensitivity of synchrotron radiation flux 
density. In case of BP1, excess in synchrotron flux density is observed for $\chi \bar{\chi}\rightarrow b \bar{b}$ annihilation when compared with the limit for SKA (for both value of $\theta$ considered). For the rest of the benchmark points, the calculated flux density
for $\chi \bar{\chi}\rightarrow b \bar{b}$ annihilation is consistent with the 
SKA bound. We further observe from Table \ref{t3} that for all BPs the flux density for synchrotron radiation
obtained for DM annihilation into $W^+ W^-$ is always larger than the bounds 
from SKA (with sensitivity $\backsim 10^{-6}$ Jy). Therefore, depending on the value aperture angle, we can expect a
formidable amount of synchrotron radiation flux produced from DM annihilations
in the region of Galactic centre that can be probed by experiments like GMRT 
and SKA. No such observation in synchrotron flux density will disfavour the 
dark matter model considered in the present work. 

We repeat our calculations (shown in Table \ref{t3}) also with the Einasto profile for dark matter densities. For demonstrative purpose we furnish the results for two benchmark points namely BP1 and BP3 in Table \ref{t5}. It can be seen from Table \ref{t5} that the flux density is reduced by around a factor of 10 when Einasto profile is used. For example for $m_{\chi}=101.5$ GeV, $\nu=$610 MHz and the aperture angle $\theta^{\prime}=10^{''}$ the flux density from the channel $\chi \bar{\chi}\rightarrow b \bar{b}$ is obtained as $9.1 \times 10^{-6}$ Jy where as the same flux with the calculated using NFW profile is $\simeq 2\times 10^{-4}$ Jy. Similar reduction of flux is also observed for $\chi \bar\chi \rightarrow W^+ W^-$ channel too.

\begin{table}[H]
\centering 
\footnotesize
\begin{tablenotes}
\centering \small
\item $B$=3 $\mu$G, Einasto density profile.
    \end{tablenotes}
\begin{tabular}{|l|l|l|l|l|l|l|l|}
\hline
BP&\hspace{2mm}$m_{\chi}$& 
\hspace{4mm}$\nu$&\hspace{2mm}$\theta^{\prime}$&\multicolumn{2}{|c|}{
observational limit}&\multicolumn{2}{|c|}{calculated flux 
density}\\\cline{5-6}\cline{7-8}
&&&&\hspace{2mm}\multirow{2}{1mm}{GMRT}  & 
\hspace{4mm}\multirow{2}{1mm}{SKA}&\hspace{8mm}\multirow{2}{1mm}{$\chi 
\bar{\chi}\rightarrow b\bar b$}  & \hspace{4mm}\multirow{2}{1mm}{$\chi 
\bar{\chi} \rightarrow W^+W^-$}\\
&&&&&&&\\
 &in GeV&in MHz&&\hspace{4mm}in Jy&\hspace{4mm}in Jy&\hspace{6mm}in 
Jy&\hspace{4mm}in Jy\\
\hline
1&\multirow{5}{*}{101.5}&325 &\multirow{3}{*}{$10^{''}$} &  0.04$\times 
10^{-3}$& \multirow{5}{*}{$10^{-6}$-$10^{-3}$} &8.0467$\times 
10^{-6}$&1.1508$\times 10^{-3}$\\ \cline{7-8} \cline{3-3} \cline{5-5}
  & &  610   & &  0.02$\times 10^{-3}$ & & 9.1059$\times 10^{-6}$&1.3023$\times 
10^{-3}$\\ \cline{7-8} \cline{3-3} \cline{5-5}
  & &  1400  & &  0.03$\times 10^{-3}$&  & 9.7919$\times 10^{-6}$&1.4004$\times 
10^{-3}$\\ \cline{7-8} \cline{3-3} \cline{4-5}
  & &325&\multirow{3}{*}{$1^{'}$} & 0.04$\times 10^{-3}$ & & 2.3854$\times 
10^{-4}$ &3.4116$\times 10^{-2}$\\ \cline{7-8} \cline{3-3} \cline{5-5}
  & & 610  & &  0.02$\times 10^{-3}$& & 2.6994$\times 10^{-4}$&3.8607$\times 
10^{-2}$\\ \cline{7-8} \cline{3-3} \cline{5-5}
  & & 1400 & &  0.03$\times 10^{-3}$&  & 2.9028$\times 10^{-4}$&4.1516$\times 
10^{-2}$\\ \cline{1-8}  
3&\multirow{5}{*}{\hspace{1mm}565}&325 &\multirow{3}{*}{$10^{''}$} &  0.04$\times 
10^{-3}$& \multirow{5}{*}{$10^{-6}$-$10^{-3}$} & 1.1057$\times 
10^{-9}$& 8.1419$\times 10^{-6}$\\ \cline{7-8} \cline{3-3} \cline{5-5}
  & &  610   & &  0.02$\times 10^{-3}$ & & 1.2513$\times 10^{-9}$& 9.2137$\times 
10^{-6}$\\ \cline{7-8} \cline{3-3} \cline{5-5}
  & &  1400  & &  0.03$\times 10^{-3}$&  & 1.3455$\times 10^{-9}$&9.9077$\times 
10^{-6}$\\ \cline{7-8} \cline{3-3} \cline{4-5}
  & &325&\multirow{3}{*}{$1^{'}$} & 0.04$\times 10^{-3}$ & &3.2778$\times 
10^{-8}$ & 2.4136$\times 10^{-4}$\\ \cline{7-8} \cline{3-3} \cline{5-5}
  & & 610  & &  0.02$\times 10^{-3}$& & 3.7093$\times 10^{-8}$& 2.7314$\times 
10^{-4}$\\ \cline{7-8} \cline{3-3} \cline{5-5}
  & & 1400 & &  0.03$\times 10^{-3}$&  &3.9888$\times 10^{-8}$& 2.9371$\times 
10^{-4}$\\ \cline{1-8}  
\end{tabular}
 \captionsetup{font=small}
\caption{ Model dependent theoretical flux densities for two chosen benchmark points (BPs, BP1 and BP3) are 
compared with the observational data are given by GMRT and SKA.}\label{t5}
\end{table}

We repeat our calculations shown in Table \ref{t3} for Jodrell Bank telescope. In Table \ref{t4} we furnish the results for all the six benchmark points considering Table \ref{t3} and for the aperture angle $\theta^{\prime}$= $4^{''}$ within the region of GC at a fixed operational frequency 408 MHz. In Table \ref{t4} we tabulated our calculated synchrotron 
flux density with chosen six BPs using a constant magnetic field value $B=$3 
$\mu$G and NFW density profile. From Table \ref{t4} it can be concluded that
synchrotron flux density for all the benchmark points are consistent with
the bound from Jodrell Bank.

\begin{table}[H]
\centering
\begin{tablenotes}
\centering
\item  $B$=3 $\mu$G, NFW density profile.
    \end{tablenotes}
\begin{tabular}{|l|l|l|l|l|l|l|l|l|l|l|}
\hline
BP&\hspace{2mm}$m_{\chi}$&\hspace{4mm}$\nu$&$\theta^{\prime}$&observational limit&\multicolumn{2}{|c|}{calculated flux density}\\\cline{5-7}
&&&&\hspace{4mm}Jodrell 
Bank&\hspace{8mm}\multirow{2}{1mm}{$\chi \bar{\chi}\rightarrow b\bar b$}  & 
\hspace{6mm}\multirow{2}{1mm}{$\chi \bar{\chi}\rightarrow W^+W^-$}\\
&&&&&&\\
&in GeV &in MHz&&\hspace{6mm}in Jy&\hspace{6mm}in Jy&\hspace{6mm}in Jy\\
\hline
1 &\hspace{1mm}{101.5}
&\multirow{5}{*}{\hspace{2mm}408}&\multirow{5}{*}{$4^{''}$}&\multirow{5}{*}{
\hspace{3mm}50$\times 10^{-3}$} &$5.3718\times10^{-5}$&$7.6826\times10^{-3}$ \\ 
\cline{8-9}
  2&\hspace{2mm}{430} & & & &$2.0262\times10^{-8}$&$8.8450\times10^{-4}$\\ \cline{8-9}
  3&\hspace{2mm}{565} & & & &$7.3814\times10^{-9}$&$5.4353\times10^{-5}$\\ \cline{8-9}
  4&\hspace{2mm}{307} & & & &$2.6014\times10^{-7}$&$5.5109\times10^{-4}$\\ \cline{8-9}
 5&\hspace{2mm}{131} & & & &$2.3407\times10^{-6}$&$7.9078\times10^{-4}$\\ \cline{8-9}
 6&\hspace{2mm}{213} & & & &$2.9646\times10^{-7}$&$3.1659\times10^{-4}$\\ \cline{1-9}
\end{tabular}
\caption{Model dependent theoretical flux densities for chosen six BPs are 
compared with the observational data as given by Jodrell Bank.}\label{t4}
\end{table}

\begin{table}[H]
\centering
\begin{tablenotes}
\centering
\item $B$=3 $\mu$G, Einasto density profile.
    \end{tablenotes}
\begin{tabular}{|l|l|l|l|l|l|l|l|l|l|l|}
\hline
BP&\hspace{2mm}$m_{\chi}$&\hspace{4mm}$\nu$&$\theta^{\prime}$&observational limit&\multicolumn{2}{|c|}{calculated flux density}\\\cline{5-7}
&&&&\hspace{4mm}Jodrell 
Bank&\hspace{8mm}\multirow{2}{1mm}{$\chi \bar{\chi}\rightarrow b\bar b$}  & 
\hspace{6mm}\multirow{2}{1mm}{$\chi \bar{\chi}\rightarrow W^+W^-$}\\
&&&&&&\\
&in GeV &in MHz&&\hspace{6mm}in Jy&\hspace{6mm}in Jy&\hspace{6mm}in Jy\\
\hline
1 &\hspace{1mm}{101.5}
&\multirow{2}{*}{\hspace{2mm}408}&\multirow{2}{*}{$4^{''}$}&\multirow{2}{*}{
\hspace{3mm}50$\times 10^{-3}$} &$1.0475\times10^{-6}$&$1.4981\times10^{-4}$ \\ 
\cline{8-9}
3 &\hspace{2mm}{565}
&&& & $1.4393\times10^{-10}$& $1.0599\times10^{-6}$ \\ 
\cline{8-9}
 \cline{1-9}
 \cline{1-9}
\end{tabular}
\caption{ Model dependent theoretical flux densities for two chosen BPs (BP1 and BP3) are 
compared with the observational data as given by Jodrell Bank.}\label{t6}
\end{table}  

We repeat our calculations for the case of Jodrell Bank telescope (Table \ref{t4}) by considering the Einasto profile for dark matter densities. In Table \ref{t6} we show the results for two benchmark points namely BP1 and BP3 with $\nu=408$ MHz and $\theta^{\prime}$= $4^{''}$ for comparision with the results given in Table \ref{t4} (using NFW profile). Here too one notices the reduction of calculated flux when Einasto profile is used. From these analyses, we conclude that the dominant dependence of flux density 
is on the dark matter mass and the corresponding annihilation cross-sections
when other parameters (such as magnetic field $B$, angular aperture 
$\theta^{\prime}$, frequency range, dark matter density profile etc.) are kept fixed.
\section{Summary and Conclusions}
In this work we explore the possibility that the synchrotron emission from the Galactic Centre region can be a dark matter indirect detection signal if detected by the ongoing and future radio telescopes such as GMRT, SKA etc. If the dark matter is accumulated in considerable numbers at a very high gravitating object such as GC, they can undergo self-annihilation and in case the final product is $e^+e^-$ pairs, these can emit synchrotron radiation in the radio wave region under the influence of the magnetic field at the GC and its vicinity. Such emissions if detected by the earthbound radio telescopes can well be a new signature for dark matter indirect detections.

With this in view, in this work we consider a particle candidate for dark matter by minimal extension of Standard Model of particle physics with a Dirac fermion and a pseudoscalar. After spontaneous symmetry breaking both the SM sector and the additional pseudoscalar acquire VEV and these two undergo mixing. The mass matrix when diagonalised give two mass eigenstates (physical states) each of which is the mixture of the two scalars, that depend on the mixing angle $\theta$ which is a parameter of the model. One of them is identified as the physical Higgs and the other is the physical pseudoscalar. The Dirac fermion is the particle dark matter candidate in this model which is connected to SM sector by Higgs portal and the pseudoscalar. We constrain our model parameters (coupling, mixing angle etc.) by theoretical bounds such as vacuum stability, perturbativity etc. as also from the experimental and collider bounds such as LHC, PLANCK limit on the dark matter relic density, XENON-1T, LUX DM direct detection bounds on dark matter nucleon scattering cross-section etc. With our fermionic dark matter thus established we calculate the annihilation cross-sections for the dark matter candidate that yield $e^+e^-$ as the end product and then compute the synchrotron flux that can be produced at the GC region. We then compare our results with the radio wave detectability for the ongoing radio telescope GMRT and the upcoming radio telescope SKA. The calculations are made with an aperture angle $\theta^{\prime}=10^{''}$ and $\theta^{\prime}=1^{'}$ in the region of GC at three fixed operational frequencies namely 325, 610 and 1400 MHz. We also compare our results with the experimental bounds on synchrotron flux (with $\theta^{\prime}=4^{''}$ and 408 MHz) obtained by Jodrell Bank telescope. We adopt two values of magnetic field; 3 $\mu$G and 6 $\mu$G. We perform the study for few chosen benchmark points and found that excess synchrotron flux can be produced near Galactic Centre due to dark matter annihilation. Therefore our model may be able to explain any excess radio fluxes if detected within the limits of our calculations, by the radio telescopes considered here as a possible dark matter indirect signature. The uncertainties in the calculations may arise from the factors such as the choice of dark matter density profile, uncertainties in the GC magnetic field etc. For the magnetic field, as mentioned, we have chosen several values $\leq$ 10 $\mu$G but obtained no significant variations of results. But if there are directional dependence in the magnetic field the uncertainties may creep in. The uncertainties due to dark matter density profile is addressed by making the calculations with two density profiles namely NFW and Einasto profiles. We find that the results differ for the two choices. Therefore suitable and proper choice of density profile is essential to reduce the errors arising out of the dark matter density profile. The main background sources are coming from astrophysical sources, cosmic rays etc. The future radio  
telescope's (SKA, uGMRT, MerrKAT) data on source counts and angular correlations may be useful  
for isolating the dark matter signals from strong backgrounds. A conservative approach is to assume
total synchrotron radiation generates from dark matter only as considered in Ref. \cite{{Chan:2016}} to limit dark matter annihilation cross-section. One can also use indirect method to separate background radiation from synchrotron radiation from dark matter. Fermi-LAT provides bounds on dark matter annihilation cross-section for gamma-rays from galactic centre. Using the limit on dark matter annihilation cross-section from Fermi-LAT, we can estimate the amount of synchrotron radiation to be obtained from dark matter only and distinguish it from the background by measuring total synchrotron flux at GC. However, this is not a precise way and subject to lot of uncertainties. If the low frequency radio telescopes like GMRT, SKA operate at the frequencies at which peak frequencies are obtained in this work, then these experiments should get a better r.m.s sensitivity. We expect that in future, radio telescopes like SKA, uGMRT, MerrKAT obtain even better sensitivities and from that we may able to impose tighter constraints on the DM properties.

\vspace{5mm}
\noindent {\bf Acknowledgments}

The authors would like to thank D. Bandyopadhyay for some useful discussions. ADB acknowledges the support from Department of Science and Technology, 
Government of India, under the grant PDF/2016/002148.

\vspace{5mm}
\noindent{\bf Appendix}

In this Appendix we furnish the expressions for the annihilation cross-sections of the fermionic dark matter candidate to the possible final states considered in the work.
\begin{equation}\label{eq:48}
\begin{aligned}
\sigma  (\bar{\chi } \chi \rightarrow \bar{f} f)=\dfrac{1}{16 \pi }N_c\big(\dfrac{m_f}{v_H}\big)^2\Big(\dfrac{s}{s-4 m_{\chi }^2}\Big)^{1/2}\Big(1-\dfrac{4 m_f^2}{s}\Big)^{3/2}
\Bigg[g^2 s \sin^2\theta\cos ^2\theta\Big[\dfrac{1}{(s-m_h^2)^2 +m_h^2 \Gamma _h^2}\\+\dfrac{1}{(s-m_{\rho}^2)^2+m_{\rho}^2 \Gamma _{\rho}^2}-
\dfrac{2((s-m_h^2)(s-m_{\rho }^2)+m_h m_{\rho }\Gamma _h \Gamma _{\rho })}{((s-m_h^2)^2+m_h^2\Gamma _h^2) ((s-m_{\rho }^2)^2+m_{\rho }^2\Gamma _{\rho }^2)}\Big]+{\Lambda ^{\prime}}^2v_{\rm{rel}}^2v_H^2(s-4 m_{\chi }^2)\\\Big[\dfrac{\cos ^4\theta}{(s-m_h^2)^2+m_h^2 \Gamma _h^2}
+\dfrac{\sin ^4\theta}{(s-m_{\rho}^2)^2+m_{\rho}^2 \Gamma _{\rho}^2}
+\dfrac{2\sin ^2\theta\cos ^2\theta((s-m_h^2)(s-m_{\rho }^2)+m_h m_{\rho }\Gamma _h \Gamma _{\rho })}{((s-m_h^2)^2+m_h^2\Gamma _h^2) ((s-m_{\rho }^2)^2+m_{\rho }^2\Gamma _{\rho }^2)}\Big]\Bigg],
\end{aligned}
\end{equation}
where $m_f$ is the mass of the fermion $f$, $N_c$ is the colour quantum number, $\Gamma _h$ and $\Gamma _{\rho }$ are the total decay widths of Higgs and scalar bosons respectively.
\begin{equation}\label{eq:49}
\begin{aligned}
\sigma  (\bar{\chi } \chi \rightarrow W^+ W^-)=\dfrac{1}{16 \pi s}\Big(\dfrac{s}{s-4 m_{\chi }^2}\Big)^{1/2}\Big(1-\dfrac{4 m_W^2}{s}\Big)^{1/2}\Big(2+\dfrac{(s-2 m_W^2)^2}{4 m_W^4}\Big)
\Bigg[2\big(\dfrac{m_W^2}{v_H}\big)^2 g^2 s \hspace{4cm} \\ \sin^2\theta\cos ^2\theta\Big[\dfrac{1}{(s-m_h^2)^2 +m_h^2 \Gamma _h^2}+\dfrac{1}{(s-m_{\rho}^2)^2+m_{\rho}^2 \Gamma _{\rho}^2}-\dfrac{2((s-m_h^2)(s-m_{\rho }^2)+m_h m_{\rho }\Gamma _h \Gamma _{\rho })}{((s-m_h^2)^2+m_h^2\Gamma _h^2) ((s-m_{\rho }^2)^2+m_{\rho }^2\Gamma _{\rho }^2)}\Big]\hspace{2.5cm}\\+2m_W^4{\Lambda ^{\prime}}^2v_{\rm{rel}}^2(s-4 m_{\chi }^2) \Big[\dfrac{\cos ^4\theta}{(s-m_h^2)^2+m_h^2 \Gamma _h^2}
+\dfrac{\sin ^4\theta}{(s-m_{\rho}^2)^2+m_{\rho}^2 \Gamma _{\rho}^2}  \hspace{8cm}\\
+\dfrac{2\sin ^2\theta\cos ^2\theta((s-m_h^2)(s-m_{\rho }^2)+m_h m_{\rho }\Gamma _h \Gamma _{\rho })}{((s-m_h^2)^2+m_h^2\Gamma _h^2) ((s-m_{\rho }^2)^2+m_{\rho }^2\Gamma _{\rho }^2)}\Big]\Bigg],\hspace{9.8cm}
\end{aligned},
\end{equation}

\begin{equation}\label{eq:50}
\begin{aligned}
\sigma  (\bar{\chi } \chi \rightarrow Z Z)=\dfrac{1}{16 \pi s}\Big(\dfrac{s}{s-4 m_{\chi }^2}\Big)^{1/2}\Big(1-\dfrac{4 m_Z^2}{s}\Big)^{1/2}\Big(2+\dfrac{(s-2 m_Z^2)^2}{4 m_Z^4}\Big)
\Bigg[\big(\dfrac{m_Z^2}{v_H}\big)^2 g^2 s \hspace{4.5cm} \\ \sin^2\theta\cos ^2\theta \Big[\dfrac{1}{(s-m_h^2)^2 +m_h^2 \Gamma _h^2}+\dfrac{1}{(s-m_{\rho}^2)^2+m_{\rho}^2 \Gamma _{\rho}^2}-\dfrac{2((s-m_h^2)(s-m_{\rho }^2)+m_h m_{\rho }\Gamma _h \Gamma _{\rho })}{((s-m_h^2)^2+m_h^2\Gamma _h^2) ((s-m_{\rho }^2)^2+m_{\rho }^2\Gamma _{\rho }^2)}\Big]\hspace{1.5cm}\\+m_Z^4{\Lambda ^{\prime}}^2v_{\rm{rel}}^2(s-4 m_{\chi }^2) \Big[\dfrac{\cos ^4\theta}{(s-m_h^2)^2+m_h^2 \Gamma _h^2}
+\dfrac{\sin ^4\theta}{(s-m_{\rho}^2)^2+m_{\rho}^2 \Gamma _{\rho}^2}  \hspace{6cm}\\
+\dfrac{2\sin ^2\theta\cos ^2\theta((s-m_h^2)(s-m_{\rho }^2)+m_h m_{\rho }\Gamma _h \Gamma _{\rho })}{((s-m_h^2)^2+m_h^2\Gamma _h^2) ((s-m_{\rho }^2)^2+m_{\rho }^2\Gamma _{\rho }^2)}\Big]\Bigg],\hspace{7.5cm}
\end{aligned}
\end{equation}
where $m_W$ and $m_Z$ are the masses of $W$ bosons and $Z$ bosons respectively. 
\begin{equation}\label{eq:51}
\begin{aligned}
\sigma  (\bar{\chi } \chi \rightarrow hh)=\dfrac{1}{16 \pi s}\Big(\dfrac{s}{s-4 m_{\chi }^2}\Big)^{1/2}\Big(1-\dfrac{4 m_h^2}{s}\Big)^{1/2}
\Bigg[\dfrac{g^2 s}{2}\Big[\dfrac{a^2 \sin ^2\theta}{(s-m_h^2)^2 +m_h^2 \Gamma _h^2}+\dfrac{b^2 \cos ^2 \theta}{(s-m_{\rho}^2)^2+m_{\rho}^2 \Gamma _{\rho}^2}\\+\dfrac{2ab\sin \theta \cos \theta((s-m_h^2)(s-m_{\rho }^2)+m_h m_{\rho }\Gamma _h \Gamma _{\rho })}{((s-m_h^2)^2+m_h^2\Gamma _h^2) ((s-m_{\rho }^2)^2+m_{\rho }^2\Gamma _{\rho }^2)}\Big]+\dfrac{{\Lambda ^{\prime}}^2v_{\rm{rel}}^2v_H^2}{2}(s-4 m_{\chi }^2) \Big[\dfrac{a^2 \cos ^2\theta}{(s-m_h^2)^2 +m_h^2 \Gamma _h^2}\\+\dfrac{b^2 \sin ^2 \theta}{(s-m_{\rho}^2)^2+m_{\rho}^2 \Gamma _{\rho}^2}+\dfrac{2ab\sin \theta \cos \theta((s-m_h^2)(s-m_{\rho }^2)+m_h m_{\rho }\Gamma _h \Gamma _{\rho })}{((s-m_h^2)^2+m_h^2\Gamma _h^2) ((s-m_{\rho }^2)^2+m_{\rho }^2\Gamma _{\rho }^2)}\Big]\Bigg],\hspace{3cm}
\end{aligned}
\end{equation}
\begin{equation}\label{eq:52}
\begin{aligned}
\sigma  (\bar{\chi } \chi \rightarrow \rho \rho)=\dfrac{1}{16 \pi s}\Big(\dfrac{s}{s-4 m_{\chi }^2}\Big)^{1/2}\Big(1-\dfrac{4 m_{\rho}^2}{s}\Big)^{1/2}
\Bigg[\dfrac{g^2 s}{2}\Big[\dfrac{c^2 \sin ^2\theta}{(s-m_h^2)^2 +m_h^2 \Gamma _h^2}+\dfrac{d^2 \cos ^2 \theta}{(s-m_{\rho}^2)^2+m_{\rho}^2 \Gamma _{\rho}^2}\\+\dfrac{2cd\sin \theta \cos \theta((s-m_h^2)(s-m_{\rho }^2)+m_h m_{\rho }\Gamma _h \Gamma _{\rho })}{((s-m_h^2)^2+m_h^2\Gamma _h^2) ((s-m_{\rho }^2)^2+m_{\rho }^2\Gamma _{\rho }^2)}\Big]+\dfrac{{\Lambda ^{\prime}}^2v_{\rm{rel}}^2v_H^2}{2}(s-4 m_{\chi }^2) \Big[\dfrac{c^2 \cos ^2\theta}{(s-m_h^2)^2 +m_h^2 \Gamma _h^2}\\+\dfrac{d^2 \sin ^2 \theta}{(s-m_{\rho}^2)^2+m_{\rho}^2 \Gamma _{\rho}^2}+\dfrac{2cd\sin \theta \cos \theta((s-m_h^2)(s-m_{\rho }^2)+m_h m_{\rho }\Gamma _h \Gamma _{\rho })}{((s-m_h^2)^2+m_h^2\Gamma _h^2) ((s-m_{\rho }^2)^2+m_{\rho }^2\Gamma _{\rho }^2)}\Big]\Bigg],\hspace{3cm}
\end{aligned}
\end{equation}
\begin{equation}\label{eq:97}
\begin{aligned}
\sigma  (\bar{\chi } \chi \rightarrow h \rho)=\dfrac{1}{16 \pi s}\Big(\dfrac{s}{s-4 m_{\chi }^2}\Big)^{1/2}\Big(1-\dfrac{({m_h+m_{\rho}})^2}{s}\Big)^{1/2} \Big(1-\dfrac{({m_h-m_{\rho}})^2}{s}\Big)^{1/2} \hspace{6cm}\\
\Bigg[\dfrac{g^2 s}{2}\Big[\dfrac{b^2 \sin ^2\theta}{(s-m_h^2)^2 +m_h^2 \Gamma _h^2} +\dfrac{c^2 \cos ^2 \theta}{(s-m_{\rho}^2)^2+m_{\rho}^2 \Gamma _{\rho}^2}+\dfrac{2bc\sin \theta \cos \theta((s-m_h^2)(s-m_{\rho }^2)+m_h m_{\rho }\Gamma _h \Gamma _{\rho })}{((s-m_h^2)^2+m_h^2\Gamma _h^2) ((s-m_{\rho }^2)^2+m_{\rho }^2\Gamma _{\rho }^2)}\Big] \hspace{3cm}\\+\dfrac{{\Lambda ^{\prime}}^2v_{\rm{rel}}^2v_H^2}{2}(s-4 m_{\chi }^2) \Big[\dfrac{b^2 \cos ^2\theta}{(s-m_h^2)^2 +m_h^2 \Gamma _h^2}+\dfrac{c^2 \sin ^2 \theta}{(s-m_{\rho}^2)^2+m_{\rho}^2 \Gamma _{\rho}^2} \hspace{9cm}\\+\dfrac{2bc\sin \theta \cos \theta((s-m_h^2)(s-m_{\rho }^2)+m_h m_{\rho }\Gamma _h \Gamma _{\rho })}{((s-m_h^2)^2+m_h^2\Gamma _h^2) ((s-m_{\rho }^2)^2+m_{\rho }^2\Gamma _{\rho }^2)}\Big]\Bigg].\hspace{10.4cm}
\end{aligned}
\end{equation}

The expressions for the couplings are given in the following
\begin{equation}\label{eq:78}
g_{\bar{\chi }\chi h }=-ig\sin \theta \gamma ^5-\Lambda^{ \prime}v_H\cos\theta,
\end{equation}
\begin{equation}\label{eq:79}
g_{\bar{\chi }\chi \rho }=-ig\cos \theta \gamma ^5+\Lambda^{ \prime}v_H\sin\theta,
\end{equation}
\begin{equation}\label{eq:80}
g_{\bar{f}fh}=\dfrac{m_f}{v_H}\cos\theta,
\end{equation}
\begin{equation}\label{eq:81}
g_{\bar{f}f\rho}=-\dfrac{m_f}{v_H}\sin\theta,
\end{equation}
\begin{equation}\label{eq:82}
g_{W^+ W^- h}=\dfrac{2 m_W^2}{v_H}\cos \theta ,
\end{equation}
\begin{equation}\label{eq:83}
g_{W^+ W^- \rho}=-\dfrac{2 m_W^2}{v_H}\sin \theta,
\end{equation}
\begin{equation}\label{eq:84}
g_{ZZh}=\dfrac{m_W^2}{v_H}\cos \theta ,
\end{equation}
\begin{equation}\label{eq:85}
g_{ZZ \rho}=-\dfrac{ m_W^2}{v_H}\sin \theta,
\end{equation}
\begin{equation}\label{eq:86}
\begin{aligned}
g_{hhh}=\frac{a}{6},\hspace{12cm}\\ {\rm{with}}\hspace{1mm}
a=-6 \lambda _H v_H\cos ^3\theta-6 \lambda _1 v_H\sin ^2\theta \cos \theta-\lambda v_{\phi }\sin ^3\theta-6 \lambda _1 v_{\phi }\sin \theta \cos ^2\theta,  \hspace{-.4cm}
\end{aligned}
\end{equation}
\begin{equation}\label{eq:87}
\begin{aligned}
g_{\rho hh}=\frac{b}{2},  \hspace{12cm}\\ {\rm{with}}\hspace{1mm}
b=6 \lambda _H v_H\sin \theta \cos ^2\theta-\lambda v_{\phi }\sin ^2\theta \cos \theta +4 \lambda _1 v_{\phi }\sin ^2\theta \cos \theta \hspace{1.8cm}\\ -4\lambda _1 v_H\sin \theta \cos ^2\theta -2 \lambda _1 v_{\phi}\cos^3\theta+2 \lambda _1 v_H\sin ^3\theta, \hspace{3cm}
\end{aligned}
\end{equation}
\begin{equation}\label{eq:88}
\begin{aligned}
g_{\rho \rho h}=\dfrac{c}{2},\hspace{12cm}\\ {\rm{with}}\hspace{1mm}
c=4 \lambda _1 v_{\phi }\sin\theta \cos ^2\theta-6 \lambda _H v_H \sin ^2\theta\cos\theta-\lambda v_{\phi }\sin \theta\cos ^2\theta-2 \lambda _1 v_{\phi}\sin ^3\theta   \hspace{-.8cm}\\+4 \lambda _1 v_H \sin ^2\theta \cos \theta-2 \lambda _1 v_H\cos ^3\theta ,  \hspace{5.5cm}
\end{aligned}
\end{equation}
\begin{equation}\label{eq:89}
\begin{aligned}
g_{\rho \rho \rho}=\dfrac{1}{6}
d, \hspace{12cm}\\{\rm{with}}\hspace{1mm}
d=6\lambda _H v_H \sin ^3\theta- \lambda v_{\phi }\cos ^3\theta-6 \lambda _1 v_{\phi} \sin^2\theta \cos \theta+6 \lambda _1 v_H\sin \theta \cos ^2\theta.
\end{aligned}
\end{equation}
{}
\end{document}